\RequirePackage{rotating}
\documentclass[a4paper,fleqn,usenatbib]{mnras}
\usepackage{newtxtext,newtxmath}
\usepackage{times}
\usepackage{rotating}
\usepackage{epsfig}
\usepackage{fancyhdr}
\usepackage{epsfig}
\usepackage{graphicx}
\usepackage{amsmath}
\usepackage{theorem}
\usepackage{amssymb}
\usepackage{latexsym}
\usepackage{epic}
\usepackage{natbib} 
\usepackage[flushleft]{threeparttable}
\usepackage{rotating}
\usepackage{tabulary}
\usepackage{tabularx}
\usepackage{pdflscape}
\usepackage{afterpage}
\usepackage{capt-of}
\usepackage{adjustbox,lipsum}
\usepackage{xcolor}
\usepackage{hyperref}
\let\la=\lesssim
\input{epsf}

\title[\textsl{NuSTAR} Observations of X-ray Faint AGN]{\textsl{NuSTAR} Observations of Four Nearby X-ray Faint AGN: Low Luminosity or Heavy Obscuration?}
\author[A. Annuar et al.]{A. Annuar$^{1,2}$\thanks{E-mail:
adlyka@ukm.edu.my (AA)}, D. M. Alexander$^{2}$, P. Gandhi$^{3}$, G. B. Lansbury$^{4,5}$, D. Asmus$^{3,6}$, 
\newauthor M. Balokovi\'{c}$^{7,8}$, D. R. Ballantyne$^{9}$, F. E. Bauer$^{10,11,12}$, P. G. Boorman$^{13,3}$,
\newauthor W. N. Brandt$^{14,15,16}$, M. Brightman$^{17}$, C.-T. J. Chen$^{18}$, A. Del Moro$^{19}$,
\newauthor  D. Farrah$^{20,21}$, F. A. Harrison$^{17}$, M. J. Koss$^{22}$, L. Lanz$^{23}$, S. Marchesi$^{24,25}$, 
\newauthor A. Masini$^{26}$, E. Nardini$^{27, 28, 2}$, C. Ricci$^{29,30,31}$, D. Stern$^{32}$, and L. Zappacosta$^{33}$
\newauthor \\
Author affiliations are listed at the end of this paper.}

\def\lsun{L$_\odot$}
\def\msun{M$_\odot$}

\begin{document}

\date{Accepted 2020 June 15. Received 2020 June 8; in original form 2020 March 23}

\pagerange{\pageref{firstpage}--\pageref{lastpage}} \pubyear{2020}

\maketitle

\label{firstpage}

\begin{abstract}
We present \textsl{NuSTAR} observations of four active galactic nuclei (AGN) located within 15 Mpc. These AGN, namely ESO 121-G6, NGC 660, NGC 3486 and NGC 5195, have observed X-ray luminosities of $L_{\rm 2-10\ keV, obs} \lesssim$ 10$^{39}$ erg s$^{-1}$, classifying them as low luminosity AGN (LLAGN). We perform broadband X-ray spectral analysis for the AGN by combining our \textsl{NuSTAR} data with \textsl{Chandra} or \textsl{XMM-Newton} observations to directly measure their column densities ($N_{\rm H}$) and infer their intrinsic power. We complement our X-ray data with archival and new high angular resolution mid-infrared (mid-IR) data for all objects, except NGC 5195. Based on our X-ray spectral analysis, we found that both ESO 121-G6 and NGC 660 are heavily obscured ($N_{\rm H}$ > 10$^{23}$ cm$^{-2}$; $L_{\rm 2-10\ keV,\ int} \sim$ 10$^{41}$ erg s$^{-1}$), and NGC 660 may be Compton-thick. We also note that the X-ray flux and spectral slope for ESO 121-G6 have significantly changed over the last decade, indicating significant changes in the obscuration and potentially accretion rate. On the other hand, NGC 3486 and NGC 5195 appear to be unobscured and just mildly obscured, respectively, with $L_{\rm 2-10\ keV,\ int} <$ 10$^{39}$ erg s$^{-1}$; i.e., genuine LLAGN. Both of the heavily obscured AGN have $L_{\rm bol} >$ 10$^{41}$ erg s$^{-1}$ and $\lambda_{\rm Edd} \gtrsim$ 10$^{-3}$, and are detected in high angular resolution mid-IR imaging, indicating the presence of obscuring dust on nuclear scale. NGC 3486 however, is undetected in high-resolution mid-IR imaging, and the current data do not provide stringent constraints on the presence or absence of obscuring nuclear dust in the AGN.
\end{abstract}



\section{Introduction}

\begin{figure*}
\centering
  \includegraphics[scale=0.433]{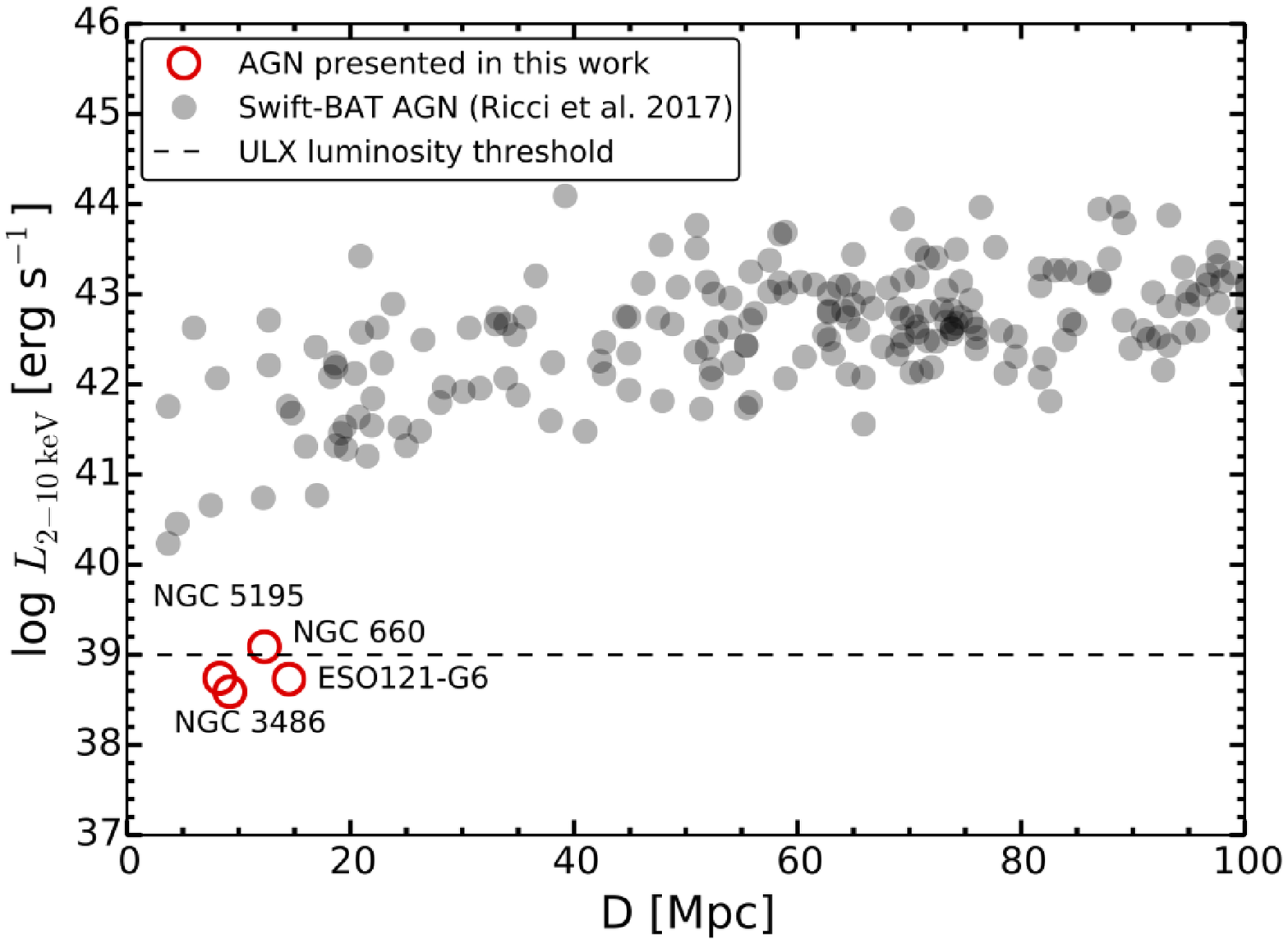}
  \includegraphics[scale=0.433]{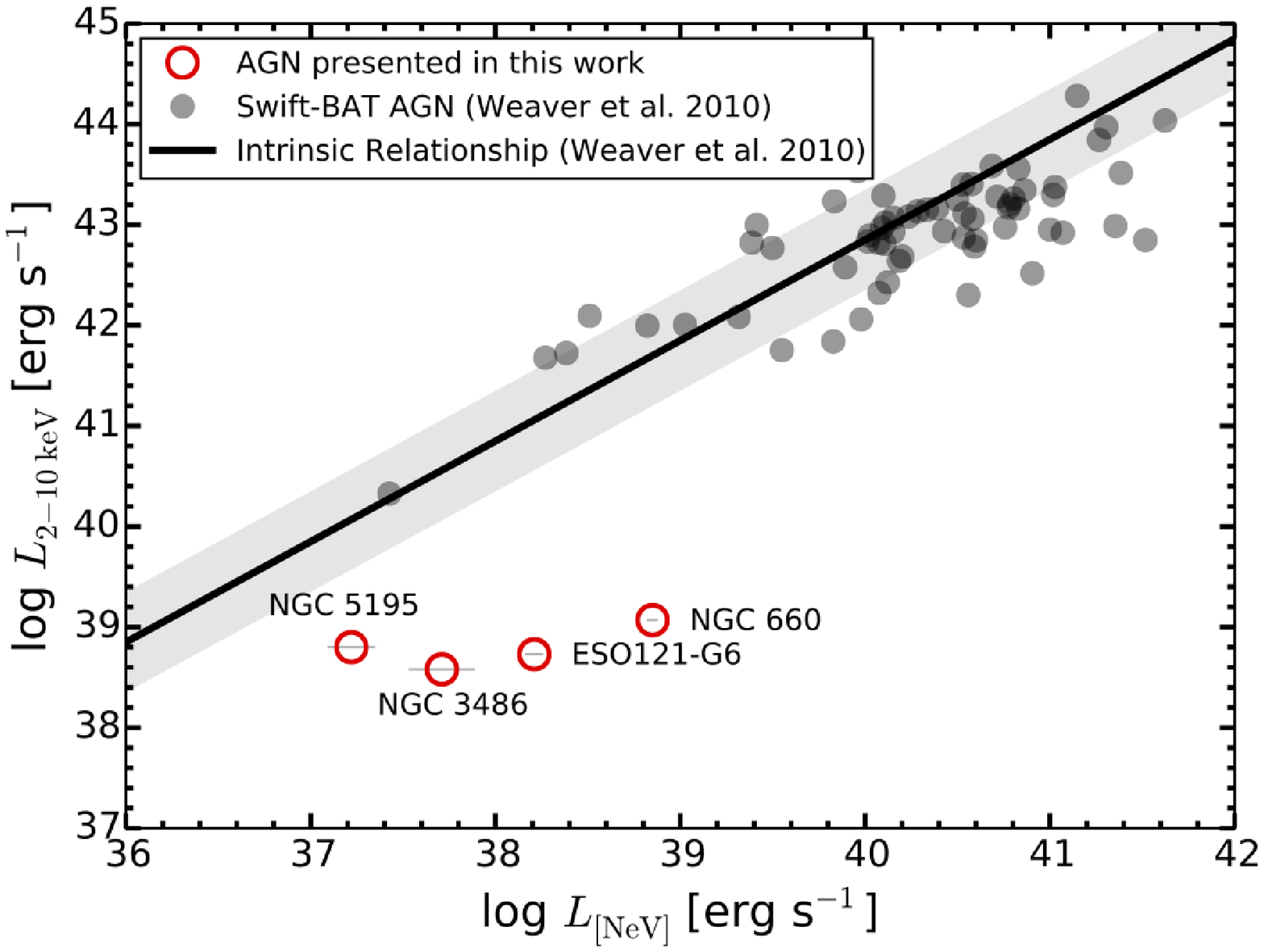}
\caption{{\it{Left}}: 2--10 keV luminosity vs. distance for the AGN presented in this work (observed luminosity; red), in comparison with the \textsl{Swift}-BAT AGN (intrinsic luminosity; grey; \citealp{Ricci15}). Our AGN are approximately an order of magnitude fainter than the limit of the \textsl{Swift}-BAT AGN. The dashed line indicates the luminosity threshold for ultraluminous X-ray sources (ULXs). {\it{Right}}: Observed 2--10 keV luminosity vs. [Ne{\sc{v}}]$\lambda$14.3$\mu$m luminosity for the AGN presented in this work (red). The [Ne{\sc{v}}] luminosity for NGC 3486 was predicted from its [O{\sc{iv}}] luminosity as it does not have a [Ne{\sc{v}}] line measurement from {\it{Spitzer}}. The solid line corresponds to the intrinsic relationship derived by \citet{Weaver10} ($\sigma \approx$ 0.5 dex) using \textsl{Swift}-BAT AGN (grey). On the basis of this simple comparison, our AGN appear to be significantly underluminous in X-rays.}
\end{figure*}

Based on the unification model of active galactic nuclei (AGN; e.g., \citealp{Antonucci93}; \citealp{Urry95}; \citealp{Bianchi12}; \citealp{Netzer15}; \citealp{AlmeidaRicci17}), the different characteristics seen in the optical spectra of Type 1 and Type 2 Seyferts are caused by the viewing angle towards the central region of the AGN. Depending on the orientation of the AGN system with respect to our line-of-sight (l.o.s.), an optically and geometrically thick region (torus) of gas and dust can obscure our direct view toward the broad-line region (BLR), resulting in the different properties that we observe for the two AGN classes. The direct identification of the broad-line emission results in a Type 1 classification, whilst the apparent absence of the BLR results in a Type 2 classification.  One of the key observational pieces of evidence supporting this theory comes from spectropolarimetry in which some Type 2 sources show broad permitted lines in their polarised spectra, consistent with that seen in Type 1 Seyfert total spectra, indicating that the nuclear regions of Type 2 Seyferts are obscured from our direct view, but can be seen if the emission is scattered into our l.o.s. (e.g., \citealp{Antonucci85}; \citealp{Capetti95}; \citealp{Kishimoto99}; \citealp{Antonucci02}).

Despite being successful in describing the physical structure for the majority of nearby AGN, there is some evidence that this model might not be valid for AGN with low luminosities ($L_{\rm bol} \lesssim$ 10$^{42}$ erg s$^{-1}$) and low accretion rates ($L_{\rm bol}$/$L_{\rm Edd} \lesssim$ 10$^{-3}$). In particular, the BLR and obscuring structure, which are probably supported by radiation pressure, are expected to collapse and disappear if the pressure drops too low (e.g., \citealp{Elitzur06}; \citealp{Hoenig07}; \citealp{Elitzur09}). There is some observational support for this basic picture (e.g., \citealp{Maoz05}; \citealp{Ho08}; \citealp{Trump11}; \citealp{Hernandez-Garcia16}; \citealp{Gonzalez-Martin17}), although due to the intrinsic faintness of the low luminosity AGN (LLAGN) emission, the current data is limited in many cases. Several studies have also predicted that LLAGN lack a standard accretion disc \citep{Shakura73}, and instead are powered by an advection-dominated accretion flow at the central region (e.g., \citealp{Narayan98}; \citealp{Quataert01}; \citealp{She18}). This is supported by observational evidence through the lack of an ultraviolet bump in the spectral energy distribution of LLAGN, which is a signature for an optically thick, geometrically thin accretion disk (e.g., \citealp{Ho99}; \citealp{Nemmen06}; \citealp{Eracleous10}). The lack of broad Fe K$\alpha$ emission in many LLAGN also supports this view as it suggests the absence (or truncation) of a standard accretion disc (e.g., \citealp{Terashima02}). 

The study of LLAGN is therefore important in our understanding of the AGN physical structure and accretion physics, as well as for building a complete census of AGN over a broad range in luminosity. The faintness of LLAGN, however, makes them challenging to study. In this paper, we define LLAGN as those with intrinsic 2--10 keV X-ray luminosity $L_{\rm 2-10\ keV,\ int} <$ 10$^{40}$ erg s$^{-1}$.

\begin{table*}
\caption{AGN and their basic properties.}
\centering
\begin{adjustbox}{width=\textwidth,totalheight=\textheight,keepaspectratio}
\begin{tabular}{lcccccccccc}
  \hline
   Name & D & Hubble & Spectral  & log $M_{\rm BH}$ & $\log{L_{[\rm OIV]}}$ &  $\log{L_{[\rm NeV]}}$ &  $\log{L_{\rm 12\mu m}}$ & $\log{L_{[\rm OIII], corr}}$ & $\log{L_{\rm IR}}$  \\
             &  [Mpc] & Type   & Class  & [\msun] & [erg s$^{-1}$] & [erg s$^{-1}$] & [erg s$^{-1}$] & [erg s$^{-1}$] & [\lsun]  \\
   (1) & (2)  & (3) & (4) & (5) & (6) & (7) & (8) & (9) & (10)  \\
   \hline
   ESO 121-G6 & 14.5  & Sc & H{\sc{ii}}  & 6.10 & 39.04& 38.21 & 40.28$^{[3]}$ &- & 9.70  \\  
   NGC 660 & 12.3  & Sa & L  & 7.35 & 39.71& 38.85& 41.18$^{[3]}$ & 40.03 & 10.49  \\
   NGC 3486 & 9.2  & Sc & S2  & 6.50$^{[1]}$ & 38.53& 37.71$^{[2]}$ & $<$ 40.10$^{[4]}$ &38.19$^{[6]}$ & 9.31  \\
   NGC 5195 & 8.3 & Irr & L  & 7.31 & 37.89& 37.22& $<$ 41.90$^{[5]}$ & 37.12 & 9.50  \\ 
   \hline
\end{tabular}
\end{adjustbox}
  \begin{tablenotes}
   \footnotesize
   \item \emph{Notes.} Column (1) Galaxy name; (2) Distance in Mpc; (3) Hubble classification of the host galaxy - S: Seyfert, L: LINER, and H{\sc{ii}}: H{\sc{ii}} region; (4) Optical spectral class on the basis of emission line diagnostic diagrams (e.g., \citealp{Kewley01}); (5) Black hole mass relative to the mass of the Sun, \msun; (6)-(9) [O{\sc{iv}}], [Ne{\sc{v}}], 12 $\mu$m continuum, and [O{\sc{iii}}] (corrected for the Balmer decrement) luminosities in units of erg s$^{-1}$, respectively; (10) total IR luminosity relative to the luminosity of the Sun, \lsun. All data are taken from \citet{Goulding09} or \citet{Goulding10}, unless indicated by additional references.
   \item \emph{References.} [1] \cite{McKernan10}; [2] Predicted from the [O{\sc{iv}}]:[Ne{\sc{v}}] luminosity relationship derived by \citet{Goulding09}; [3] This work; [4] High angular resolution 12$\mu$m luminosity from \cite{Asmus14}; [5] WISE 12$\mu$m luminosity; [6] \cite{Ho97}.
   \end{tablenotes}
\end{table*} 

The observed luminosity of an AGN can also mislead our interpretation of the nature of the source. AGN can appear to be of a low luminosity, when in fact they are deeply buried from our view by the dusty torus, or larger scale obscuration, suppressing the observed emission. Many studies have shown that the majority of AGN accretion occurs in the obscured phase, in which the central engine is hidden from our view by dust/gas with column densities of $N_{\rm H}$ $\geq$ 10$^{22}$ cm$^{-2}$ (see recent review by \citealp{HickoxAlexander18}). This is also evident from the spectral shape of the cosmic X-ray background (CXB) radiation, in which a significant population of obscured AGN is required to account for the high energy peak of the CXB spectrum (e.g., \citealp{Setti89}; \citealp{Gilli07}; \citealp{Gandhi07}; \citealp{Treister09}; \citealp{DraperBallantyne10}; \citealp{Ueda14}; \citealp{Akylas12}; \citealp{Comastri15}). Many AGN population studies support the above works which indeed show that obscured AGN dominate the overall AGN population in the universe (e.g., \citealp{Risaliti99}; \citealp{Alexander01}; \citealp{Panessa06}; \citealp{Akylas09}; \citealp{BN11}; \citealp{Ajello12}; \citealp{Aird15}; \citealp{Buchner15}; \citealp{Ricci15}). Obscured AGN, however, can be very challenging to identify, especially those in which the obscuring column density exceeds the Compton-thick (CT) threshold ($N_{\rm H}$ $\gtrsim$ 1.5 $\times$ 10$^{24}$ cm$^{-2}$). Unambiguous identification of these sources requires high quality broadband X-ray spectral analysis to properly characterise their spectra and directly measure the $N_{\rm H}$ value of the obscuring material. This can be particularly challenging to achieve for distant sources and intrinsically low luminosity obscured AGN as they often require deep X-ray observations in order to gain sufficient counts for detailed X-ray spectral analyses.

Over the last few years, we started a program to study the X-ray properties of a complete sample of AGN within $D \leq$ 15 Mpc, identified on the basis of the high ionisation [Ne{\sc{v}}]$\lambda$14.3$\mu$m emission line (ionisation energy = 97.1 eV) detection \citep{Goulding09}, to form the most complete census of the CT AGN population and the $N_{\rm H}$ distribution of AGN in the local universe. Our aim is to directly measure the $N_{\rm{H}}$ values for each AGN by performing broadband X-ray spectroscopy (over $\sim$2 orders of magnitude in energy range) using data from multiple focusing X-ray observatories, primarily the \textsl{Nuclear Spectroscopic Telescope Array} (\textsl{NuSTAR}; \citealp{Harrison13}), in combination with \textsl{Chandra} and \textsl{XMM-Newton}. The results of the first two sources in the sample observed by \textsl{NuSTAR} as part of our program; i.e., NGC 5643 and NGC 1448, are presented in \cite{Annuar15} and \cite{Annuar17}, respectively. For both of these sources, we unambiguously identified the AGN as a CT AGN. In this paper, we present new \textsl{NuSTAR} observations and direct column density measurements for a further four AGN within the sample, namely, ESO 121-G6, NGC 660, NGC 3486 and NGC 5195. The observed X-ray luminosities of these AGN are of the order of 10$^{39}$ erg s$^{-1}$, comparable with the luminosity threshold for ultraluminous X-ray sources (ULXs; refer to \citealp{Kaaret17} for a recent review on ULXs; see the left panel of Figure 1).\footnote{Although originally considered candidates for intermediate mass black holes or stellar-mass black holes undergoing super-Eddington accretion (e.g., \citealp{Orosz03}), many ULXs have now been shown to be neutron stars undergoing super-Eddington accretion (e.g., \citealp{Walton18}).} The observed X-ray luminosities of our sources suggest that they are LLAGN. However, a comparison of their observed X-ray luminosities to their [Ne{\sc{v}}]$\lambda$14.3$\mu$m emission line luminosities suggests that they are underluminous in X-rays when compared to that found for typical AGN (see the right panel of Figure 1). This suggests that the X-ray emission in these AGN may be heavily obscured. 

In this paper, we perform broadband X-ray spectral analysis and explore the mid-infrared (mid-IR) properties of the AGN to investigate their nature; i.e., to explore whether they are indeed buried AGN as suggested by their [Ne{\sc{v}}] luminosity, or whether they are intrinsically LLAGN. The characterisation of these AGN is important in allowing us to build a complete census of the nearby AGN population over a broad range of obscuration and luminosities, and to help us further test AGN physical models. We describe each source, and detail their mid-IR and X-ray observations in Section 2. We also present the broadband spectral modelling and results in Section 2. In Section 3, we conduct additional X-ray and mid-IR analyses to further investigate the nature of the AGN, and discuss their properties. Finally in Section 4, we summarise our conclusions and provide brief details on our future work.

\section{Observations and Spectral Analysis}

\begin{table*}
\begin{center}
\caption{Log of X-ray observations used in this work.}
\begin{tabular}{lcccccccc}
  \hline \hline
   Name & R.A. & Dec. & Observatories & ObsID & Date & Energy Band & Net Exposure Time & Net Count Rate  \\
             &   &  &  &    &  &[keV] & [ks] & [10$^{-3}$ cts s$^{-1}$]  \\
   (1) & (2)  & (3) & (4) & (5) & (6) & (7) & (8) & (9) \\
   \hline
      ESO 121-G6 & 6:07:29.86 & $-$61:48:27.61 &\textsl{XMM-Newton} & 0403072201  & 2007-01-20 & 0.5--10 & 11.5 & 2.49 $\pm$ 0.77    \\  
  &  &  & \textsl{Chandra} & 19523 & 2017-12-23 & 0.5--8 & 32.1 & 31.15 $\pm$ 1.29   \\  
   			& & & & 20892 & 2017-12-24 & 0.5--8 & 22.0 & 30.36 $\pm$ 1.54 \\
   		      & & & \textsl{NuSTAR} & 60201063002  & 2017-12-25 & 3--50 & 98.6 & 48.41 $\pm$ 0.75 \\
   NGC 660 & 01:43:02.32 & 13:38:44.90 & \textsl{Chandra} & 15333  & 2012-12-18 & 0.5--8 & 22.8 & 4.03 $\pm$ 0.42  \\  
  		   & & & & 15587& 2012-11-20 & 0.5--8 & 27.7  & 4.24 $\pm$ 0.39  \\
 		   & & &&18352& 2015-08-26 & 0.5--8 & 10.0  & 3.29 $\pm$ 0.57  \\
   		    & & & \textsl{NuSTAR} & 60101102002 & 2015-08-23& 3--30  & 112 & 1.41 $\pm$ 0.14 \\
   NGC 3486 & 11:00:23.87 & 28:58:30.49 & \textsl{XMM-Newton} & 0112550101 & 2001-05-09 & 0.5--10 & 9.3 & 20.79 $\pm$ 0.30 \\  
   		    & & & \textsl{NuSTAR} &  60001150002 & 2015-01-26 & 3--24 & 57.8  & $<$0.74   \\
   NGC 5195 & 13:29:59.41 & 47:15:57.29  & \textsl{Chandra} & 19522 & 2017-03-17 &0.5--8 & 37.8 & 8.21 $\pm$ 0.51   \\  
   		    & & & \textsl{NuSTAR} & 60201062002 & 2017-03-16 & 3--24 & 94.3  & 0.25 $\pm$ 0.08 \\
   		    & & &  				& 60201062003 & 2017-03-17 & 3--24 &  326  & 0.50 $\pm$ 0.05 \\
   \hline
\end{tabular}
\end{center} 
  \begin{tablenotes}
   \footnotesize
   \item \emph{Notes.} (1) Galaxy name; (2)-(3) AGN position that were used to extract the spectra; (4) observatory; (5) observation identification number; (6) observation UT start date; (7) energy band in keV; (8) the net (clean) exposure time in ks; (9) net count rate within the extraction region in the given energy band in units of 10$^{-3}$ s$^{-1}$. The net exposure times and count rates for \textsl{NuSTAR} and \textsl{XMM-Newton} are the total values from the FPMs, and EPIC cameras, respectively.    
   \end{tablenotes}  
\end{table*}

In this section, we describe each AGN target, their X-ray observations, the data reduction procedures adopted in this work, and the X-ray spectral analysis of each AGN. In addition, we also detail the high spatial resolution mid-IR observations for the AGN. In Table 1 and 2, we present a summary of the AGN basic properties and their X-ray observations, respectively. We summarise the main results of our spectral analysis in Table 3.


\textsl{NuSTAR}, launched on June 13th, 2012, is the first focusing high energy X-ray observatory in orbit \citep{Harrison13}. The energy range over which it is sensitive (i.e., 3--79 keV) provides excellent coverage for detecting the characteristic signatures of obscured AGN, such as the photoelectric cut-off at $E <$ 10 keV, the fluorescence Fe K$\alpha$ line emitted at $\sim$6.4 keV, and the Compton hump at $\sim$30 keV. In addition, it also provides a $\sim$100$\times$ improvement in sensitivity and over an order of magnitude improvement in angular resolution (18$\arcsec$ for full width half maximum, FWHM; 58$\arcsec$ for half power diameter) with respect to the previous generation of high energy X-ray observatories such as \textsl{INTEGRAL} and \textsl{Swift}-BAT. These advantages make \textsl{NuSTAR} an ideal instrument to identify obscured and relatively low luminosity AGN, which may have been missed by previous high energy X-ray observatories. 

We processed the {\emph{NuSTAR}} data for our sources with the \textsl{NuSTAR} Data Analysis Software ({\sc{nustardas}}) v1.4.1 within {\sc{heasoft}} v6.15.1 with CALDB v20191219. The {\sc{nupipeline}} v0.4.3 script was used to produce the calibrated and cleaned event files using standard filter flags. We extracted the spectra and response files from each of the \textsl{NuSTAR} focal plane modules (FPM A and FPM B), using the {\sc{nuproducts}} v0.2.5 task.\footnote{Further details on the \textsl{NuSTAR} data analysis procedure can be found at https://heasarc.gsfc.nasa.gov/docs/nustar/analysis/nustar\_swguide.pdf.} The spectra and response files from each of the \textsl{NuSTAR} FPMs were combined together using the {\sc{addascaspec}} script to increase the overall signal-to-noise ratio of the data.\footnote{More details on the {\sc{addascaspec}} script can be found at https://heasarc.gsfc.nasa.gov/docs/asca/adspecinfo.html.} In addition to the spectral extraction, we also combined the \textsl{NuSTAR} event files from the two FPMs using {\sc{xselect}} to produce the total event file.\footnote{The {\sc{xselect}} user guide can be found at https://heasarc.gsfc.nasa.gov/ftools/xselect/xselect.html.} The total count images at different energy bands were then produced from the resultant event files using the {\sc{dmcopy}} task in {\sc{ciao}} (see below).

In addition to \textsl{NuSTAR}, we also used new and archival low energy X-ray observations from \textsl{Chandra} and \textsl{XMM-Newton} to facilitate our X-ray spectral analysis of the AGN at low energies ($E \lesssim$ 3 keV), where \textsl{NuSTAR} is not sensitive. The higher spatial resolution provided by \textsl{Chandra} is crucial in helping us to reliably account for contaminating emission to the \textsl{NuSTAR} spectrum from off-nuclear X-ray sources. The \textsl{Chandra} data were reprocessed to create event files with updated calibration modifications using the {\sc{ciao}} v4.6 pipeline \citep{Fruscione06} following standard procedures. We then used the {\sc{dmcopy}} task to produce X-ray images of each source in different energy bands, and extracted the source spectra using the {\sc{specextract}} task in {\sc{ciao}}. For \textsl{XMM-Newton}, we analysed the Pipeline Processing System (PPS) data products using the Science Analysis Software (SAS v13.5.0) with the standard filter flags. Background flares were excised from the data by visually examining the source light curves, and the X-ray spectra from the three EPIC cameras were then extracted using the {\sc{evselect}} task in SAS. The spectra extracted for the EPIC MOS1 and MOS2 cameras were combined using the {\sc{epicspeccombine}} task in SAS. 

We performed our spectral analysis using {\sc{xspec}} v12.8.2. We included a fixed Galactic absorption component for each source \citep{Kalberla05} using the {\sc{xspec}} model ``{\sc{phabs}}" in all spectral fits, and assumed solar abundances for all models. Due to the modest quality of our data, we also fixed the cross-calibration uncertainties of each observatory with respect to \textsl{NuSTAR} to the values found by \citet{Madsen15} using a constant parameter, \textsl{C}, unless stated otherwise. Given the non-negligible contribution of background to the weak source flux in most cases, particularly in the \textsl{NuSTAR} and \textsl{XMM-Newton} data, we binned our spectra to a minimum of 5 counts per bin for the \textsl{NuSTAR} and \textsl{XMM-Newton} data, and 1 count per bin for the \textsl{Chandra} data using the {\sc{grppha}} task in {\sc{heasoft}}, except for ESO 121-G6 (see Section 3.1).\footnote{Further details on the {\sc{grppha}} task can be found at https://heasarc.gsfc.nasa.gov/docs/journal/grppha4.html.} We then optimised the fitting parameters using the Poisson C-statistic \citep{Cash79} for all AGN with the exception for ESO 121-G6, which was fitted using the chi-squared ($\chi^{2}$) statistic. All errors are quoted at 90$\%$ confidence. 
 
\subsection{ESO 121-G6}

ESO 121-G6 is a highly inclined ($i =$ 90$^{\circ}$){\footnote{The host galaxy inclination was obtained from the HyperLeda website (http://leda.univ-lyon1.fr/).}} galaxy located at a distance of 14.5 Mpc. The AGN in the galaxy was discovered in 2009 using the [Ne{\sc{v}}] line detection \citep{Goulding09}. The source lacks sensitive nuclear optical spectroscopy, and is therefore unclassified at optical wavelengths. However, due to the edge-on inclination of the galaxy along our l.o.s., the optical emission from the AGN is expected to be severely absorbed by the host galaxy. Prior to our study, this galaxy had only been observed in X-rays by \textsl{XMM-Newton} in 2007 for 15 ks (ObsID 0403072201), in which the data show a weak point source within 5$\arcsec$ of the 2MASS position of the galaxy (RA $=$ 06:07:29.86, Dec. $= -$61:48:27.3). The average net count rate measured from the three EPIC cameras is CR$_{0.5-10}$ $\sim$ 1.3 $\times$ 10$^{-3}$ counts s$^{-1}$, corresponding to $f_{\rm 0.5-10} \sim $ 3.8 $\times$ 10$^{-15}$ erg s$^{-1}$ cm$^{-2}$, assuming a simple power-law model with Galactic absorption ($\Gamma =$ 1.74$^{+1.14}_{-1.26}$). The observed 2--10 keV luminosity of the AGN measured from these data, $L_{\rm 2-10\ keV,\ obs} \sim $ 5.4 $\times$ 10$^{38}$ erg s$^{-1}$, is significantly lower than expected from the [Ne{\sc{v}}] luminosity, suggesting heavy obscuration of the nuclear source (see Figure 1). This is also the case when we compare the X-ray luminosity with the 12$\mu$m luminosity of the AGN measured at high spatial resolution (see Section 2.1.1, Section 3 and Figure 9), providing further evidence that the AGN is heavily obscured at X-ray energies, and is possibly CT.

\subsubsection{High Spatial Resolution Mid-IR Observation}

We observed ESO 121-G6 at mid-IR wavelengths at high spatial resolution in 2010 using the Thermal-Region Camera Spectrograph (T-ReCS; field of view 28$\farcs$8 $\times$ 21$\farcs$6; 0.09 arcsec pixel$^{-1}$; \citealp{Telesco98}), mounted on the Gemini-South telescope. The observations were carried out on 2010-10-25 (Program ID GS-2010B-Q-3; PI F. Bauer) for $\approx$319 s on-source time using the $N$-band filter ($\lambda =$ 7.4--13.4 $\mu$m) in parallel chop and nod mode. We reduced the data using the {\sc{midir}} pipeline in {\sc{iraf}} provided by the Gemini Observatory, and performed the image analysis using the {\sc{idl}} package {\sc{mirphot}}, following \citet{Asmus14}. We detected a compact core on top of faint extended emission tracing the host morphology (see Figure 2). We estimated the unresolved, nuclear flux by subtracting a manually scaled point source from the image, leaving a flat residual as judged by eye. Owing to this, the uncertainty on the nuclear flux is relatively large. The flux we measured is 3 $\pm$ 2 mJy at 12$\mu$m, corresponding to a luminosity of $L_{\rm 12\mu m} =$ (1.9 $\pm$ 1.2) $\times$ 10$^{40}$ erg s$^{-1}$. 

\begin{figure}
\centering
  \includegraphics[scale=0.33]{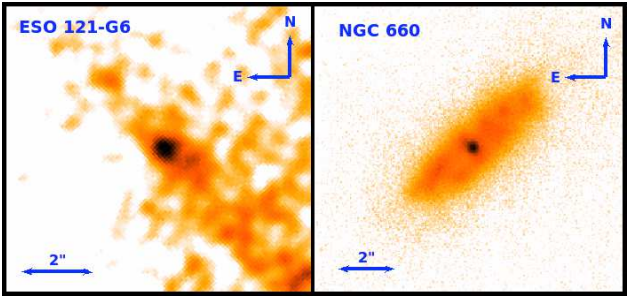}
\caption{Gemini-South T-ReCS ($\lambda_{\rm mean} = 10.3 \mu$m) and VLT VISIR mid-IR ($\lambda_{\rm mean} = 8.7 \mu$m) images of ESO 121-G6 (left) and NGC 660 (right), respectively. The ESO 121-G6 image has been smoothed with a Gaussian function of radius 3 pixels (0.3$\arcsec$) for visual clarity.}
\end{figure}

\subsubsection{X-ray Observations $\&$ Data Extraction}

ESO 121-G6 was observed at X-ray wavelength with \textsl{NuSTAR} on 2017-12-25 with an on-source exposure time of $t_{\rm exp} =$ 50.1 ks (ObsID 60201063002). Our \textsl{NuSTAR} observation was assisted by contemporaneous \textsl{Chandra} observations (ObsID 19523 and 20892; $t_{\rm exp} =$ 32.6 ks and 22.3 ks, respectively), which provided lower energy X-ray data as well as a higher angular resolution X-ray image of the galaxy. The \textsl{Chandra} data revealed two point sources within $\sim$3$\arcsec$ of the 2MASS position of the galaxy in the broad \textsl{Chandra} band of 0.5--8 keV. This is the first time that the two sources are resolved and detected in X-ray. We used the {\sc{wavdetect}} tool within {\sc{ciao}} to determine the centroid position of the two central sources in this energy band, setting the threshold parameter to 1 $\times$ 10$^{-7}$. One of the \textsl{Chandra} sources is detected at 2-8 keV, and is located at a position of RA $=$ 6:07:29.86, and Dec. $=$ $-$61:48:27.61, with formal wavdetect errors of 0$\farcs$03 and 0$\farcs$01, respectively.\footnote{The \textsl{Chandra} positional errors quoted represent 1-$\sigma$ statistical errors estimated by {\sc{wavdetect}}. These positional errors do not take into account of additional uncertainties in the source position such as the signal-to-noise ratio, off-axis angle, and astrometric uncertainties. Hence, the true uncertainty on the X-ray source position is likely to be larger than those quoted (i.e., of the order of 0$\farcs$3--0$\farcs$6; see Section 3.4.1 of \citealp{Alexander03}). We therefore assume a positional uncertainty of 0$\farcs$6 for the Chandra sources.} This is consistent with the 2MASS position of the galaxy within 0.5$\arcsec$. The other point source was detected at RA $=$ 6:07:30.12, and Dec. $=$ $-$61:48:30.21, with formal wavdetect errors of 0$\farcs$21 and 0$\farcs$09 (see footnote 7), respectively, and is undetected by \textsl{Chandra} at 2--8 keV. We therefore assumed that the former is the AGN, and used this position when extracting the X-ray spectra of the AGN. In Figure 3, we show the combined \textsl{Chandra} RGB image of ESO 121-G6. 

A source consistent with the \textsl{Chandra} position of the AGN was detected in both of the \textsl{NuSTAR} observations, with significant counts up to $\sim$50 keV. The combined \textsl{NuSTAR} RGB image of the source is shown in Figure 3. The count rate detected in the combined \textsl{NuSTAR} data in the 3--50 keV band is 4.84 $\times$ 10$^{-2}$ counts s$^{-1}$. We extracted the \textsl{NuSTAR} spectra of the AGN using a 50$\arcsec$-radius circular region, corresponding to $\sim$70$\%$ encircled energy fraction. The \textsl{Chandra} spectrum was also extracted using the same size region to match the \textsl{NuSTAR} extraction region. The background for both the \textsl{Chandra} and \textsl{NuSTAR} spectra were extracted using a circular region of 100$\arcsec$-radius from a source-free area. 

\begin{figure*}
\centering
\includegraphics[scale=0.38]{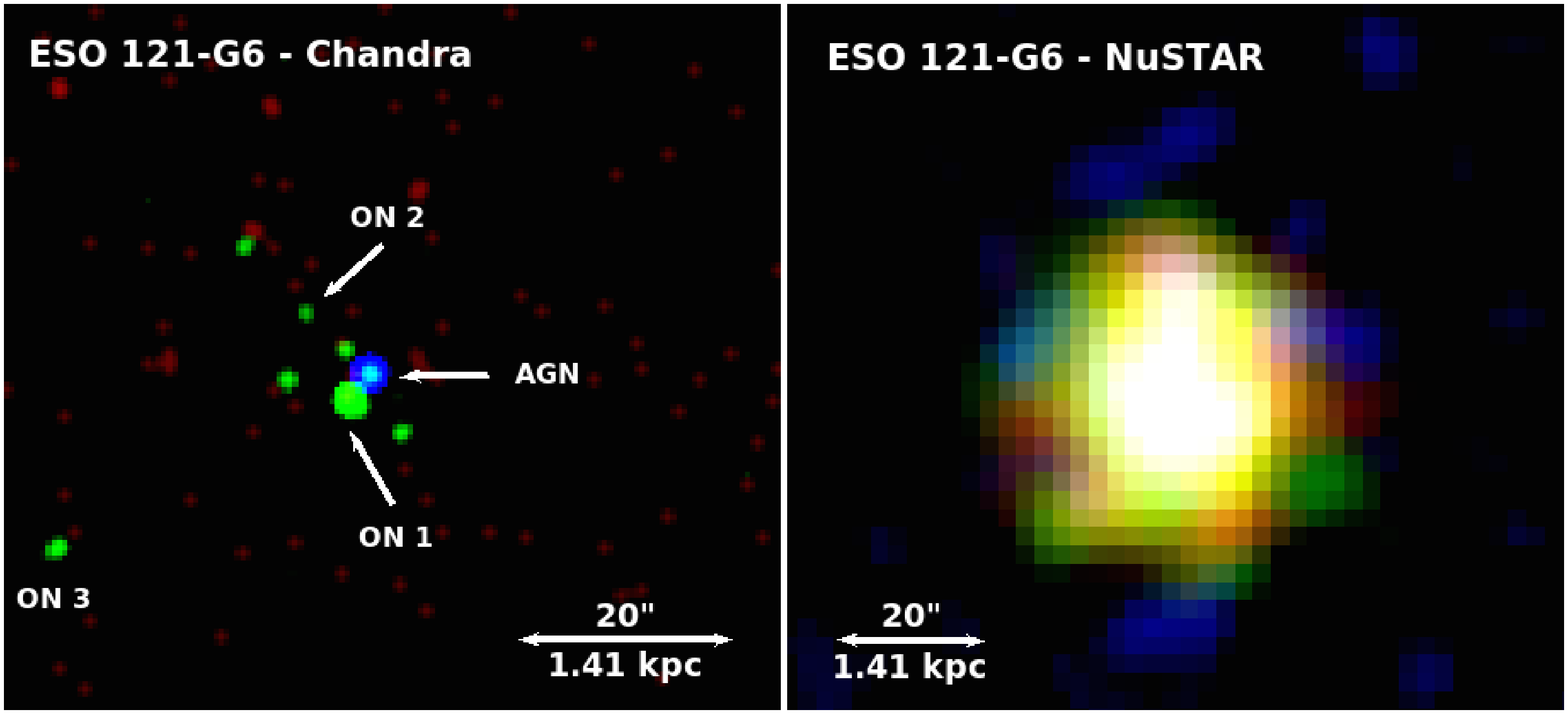} \\ \vspace{+1em}
\includegraphics[scale=0.35]{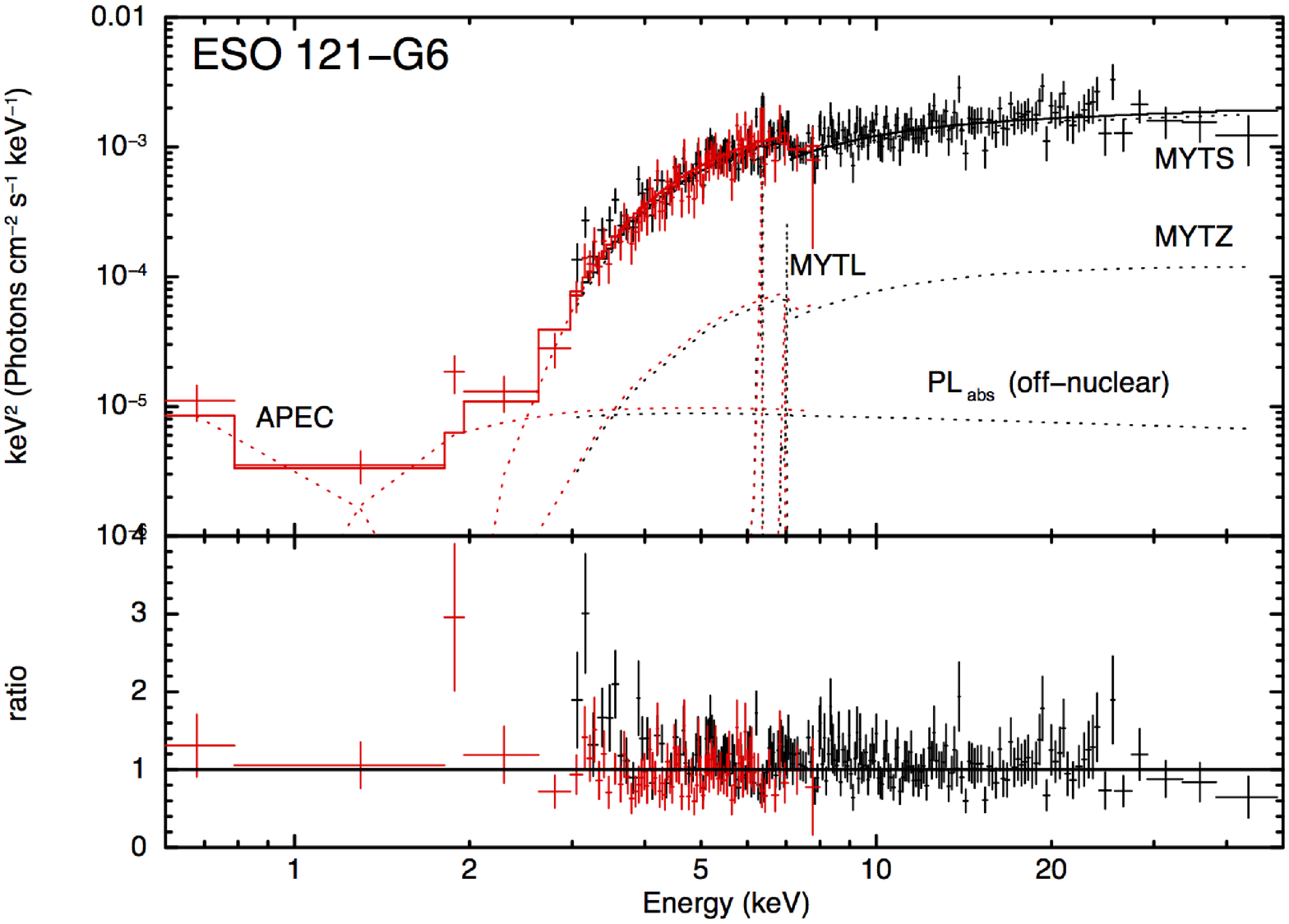}
\includegraphics[scale=0.35]{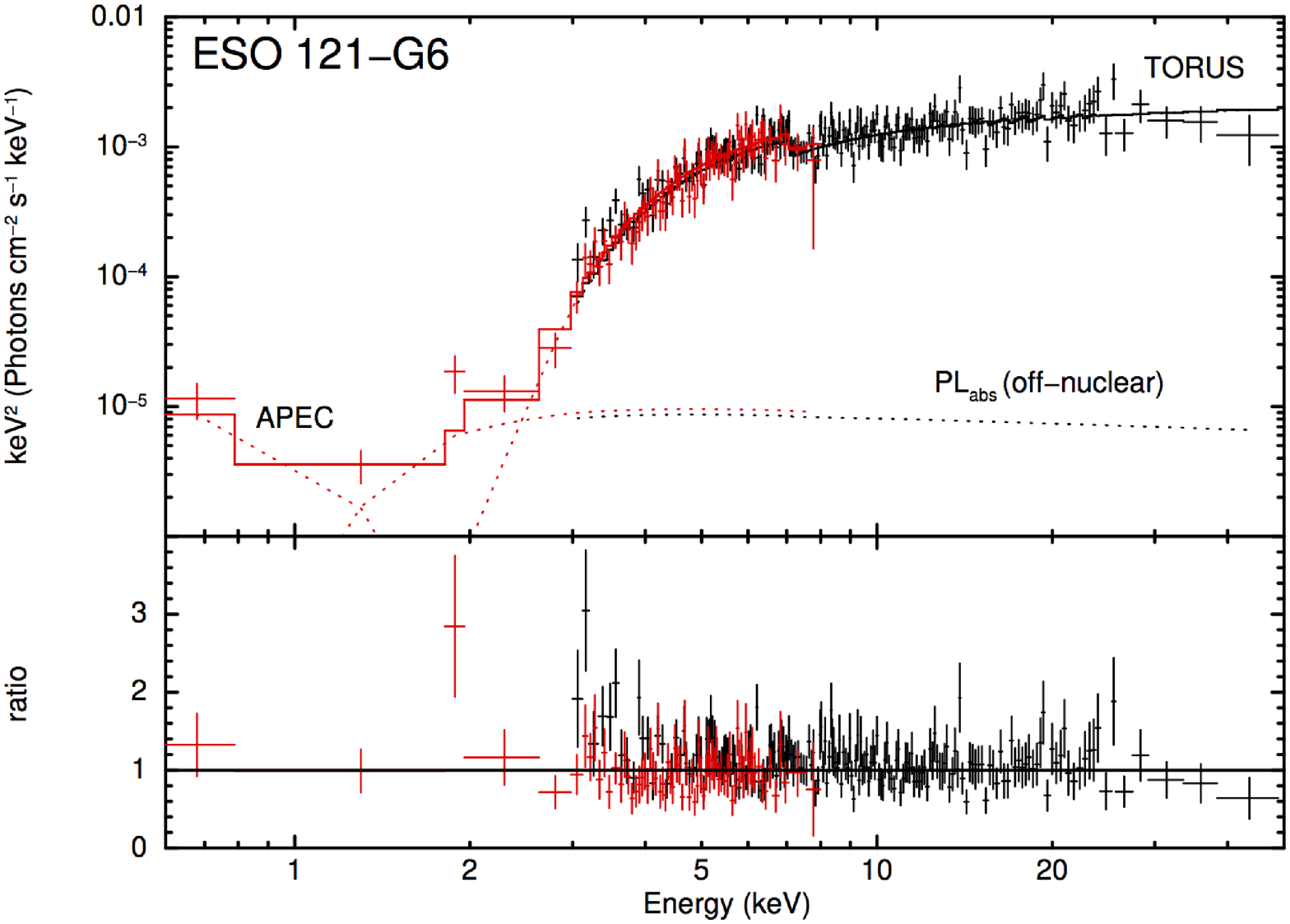}
\caption[[\textsl{Chandra} and \textsl{NuSTAR} images of ESO 121-G6 and the broadband X-ray spectrum]{\footnotesize \emph{Top}: \textsl{Chandra} and \textsl{NuSTAR} RGB images of ESO 121-G6 (\textsl{Chandra} - Red: 0.5--1 keV, Green: 1--2 keV, Blue: 2--8 keV; \textsl{NuSTAR} - Red: 3--8 keV, Green: 8--24 keV, Blue: 24--50 keV). The off-nuclear sources which were detected within the 50$\arcsec$-radius extraction region in \textsl{Chandra} are labelled as ON 1, ON 2 and ON 3. The images are smoothed with a Gaussian function of radius 3 pixels, corresponding to 1.5$\arcsec$ and and 7.4$\arcsec$ for \textsl{Chandra} and \textsl{NuSTAR}, respectively. North is up and east is to the left in all images. \emph{Bottom}: Best-fitting MY{\sc{torus}} (left) and {\sc{torus}} (right) models to the combined \textsl{NuSTAR} (black) and \textsl{Chandra} (red) data of ESO 121-G6. The data have been rebinned to a minimum of 3$\sigma$ significance with a maximum of 500 bins for visual clarity. The top panel shows the data and unfolded model in $E^{2}F_{E}$ units, whilst the bottom panel shows the ratio between the data and the folded model. The spectra were fitted using the MY{\sc{torus}} \citep{MY09} and {\sc{torus}} \citep{BN11} models to simulate the obscuring torus, an {\sc{apec}} component, a scattered power-law component to model the emission at the softest energies, as well as a power-law component to model the off-nuclear source located north-west of the AGN (ON 1). ON 2 and ON 3 were not included in the spectral fitting as they were only weakly detected, and their contributions are insignificant. The model components fitted to the data are shown as dotted curves, and the combined model fit is shown as solid curves.}
\end{figure*}

\subsubsection{X-ray Spectral Fitting}

To reliably model the X-ray spectra of the AGN in ESO 121-G6, we need to account for contributions from the off-nuclear point source detected within the extraction region of the \textsl{Chandra} data, located South-East of the AGN (labelled as ON 1 in Figure 3). We extracted the spectrum of this off-nuclear source using a small 1.5$\arcsec$-radius circular region. The background flux was extracted using a 20$\arcsec$-radius circular aperture from a source-free region. We note that there are a further two faint sources detected within the extraction region of the AGN in the 0.5--8 keV energy band (labelled as ON 2 $\&$ ON 3 in Figure 3); however, we did not include them in our spectral analysis of the AGN as they are only weakly detected, and their contributions are insignificant. We modelled the X-ray spectrum of the brightest off-nuclear source (ON 1) extracted from \textsl{Chandra} between 0.5--8 keV using a simple power-law model, absorbed by the host galaxy absorption ({\sc{tbabs}}) and Galactic column, $N_{\rm{H}}^{\rm{Gal}}$ $=$ 4.06 $\times$ 10$^{20}$ cm$^{-2}$ \citep{Kalberla05}. We assumed that the source is located within ESO 121-G6 ($z =$ 0.004039). The best-fitting photon index and host galaxy absorption measured by the model are $\Gamma$ $=$ 2.15$^{+1.01}_{-0.81}$ and $N_{\rm{H}}$ $=$ 1.23$^{+1.03}_{-0.92}$ $\times$ 10$^{22}$ cm$^{-2}$, respectively (C-stat/d.o.f $=$ 28/25). The model also inferred a 0.5--8 keV intrinsic luminosity of $L_{\rm 0.5-8\ keV,\ int} =$ 7.70$^{+1.74}_{-1.52}$ $\times$ 10$^{38}$ erg s$^{-1}$, below the luminosity threshold for a ULX, suggesting that the off-nuclear source is likely to be an X-ray binary. The measured luminosity is $\sim$2 orders of magnitude fainter than the observed luminosity measured for the AGN ($L_{\rm 0.5-8\ keV,\ obs,\ AGN} \sim$ 2.6 $\times$ 10$^{40}$ erg s$^{-1}$). 

We then proceeded to analyse the X-ray spectra of the AGN. Given the relatively high count rates for the AGN in both the \textsl{Chandra} and \textsl{NuSTAR} data, we binned the spectra to a minimum of 20 counts per bin and optimise our fitting using the chi-squared statistic. We first fitted the \textsl{Chandra} and \textsl{NuSTAR} spectra of the AGN simultaneously between 3--50 keV using a simple power-law model, absorbed by the Galactic column. The fit was poor (reduced $\chi^{2} \sim$ 3) due to significant excess between $\sim$5--10 keV, likely to be associated with spectral bump caused by obscuration. The best-fit photon index measured from this model is relatively flat $\Gamma_{3-50} \approx$ 0.95, suggesting severe absorption of the AGN flux along our l.o.s. If we measured the photon index between 2--10 keV using the same model, we obtained $\Gamma_{2-10\ keV} =$ 0.16 $\pm$ 0.07. This is significantly flatter than that measured for the archival \textsl{XMM-Newton} data in the same band; i.e., $\Gamma_{2-10\ keV}$ = 1.74$^{+1.14}_{-1.26}$, even after accounting for the large statistical uncertainties.

We then modelled the broadband X-ray spectrum of the AGN in ESO 121-G6 between 0.5--50 keV using physically motivated torus models by \citet{MY09} and \citet{BN11}, called the MY{\sc{torus}} and {\sc{torus}} models, respectively. These models were produced using a Monte Carlo approach to simulate obscuring gas and dust with different geometries. The main difference between the two models is the adopted torus geometry. The MY{\sc{torus}} model assumes a toroidal absorber geometry, and the {\sc{torus}} model simulates a spherical torus with a biconical cut out. Whilst the l.o.s. column density for the MY{\sc{torus}} model depends on the inclination angle, this is not the case for the {\sc{torus}} model. The {\sc{torus}} model can measure the column density up to $N_{\rm{H}}$ $=$ 10$^{26}$ cm$^{-2}$, but the MY{\sc{torus}} model only allows a measurement up to $N_{\rm{H}}$ $=$ 10$^{25}$ cm$^{-2}$. However, the direct ({\sc{mytz}}), scattered ({\sc{myts}}) and line components ({\sc{mytl}}) of the MY{\sc{torus}} model can be disentangled from each other, allowing more freedom in modelling of the data. We note that an improved and more complex version of {\sc{torus}} has been developed by \cite{Balokovic18}, which would allow for more accurate and detailed modelling of the AGN. However, due to the larger number of parameters and limited number of X-ray counts for our sources, we do not use this model in our paper.

For both models, we fitted the spectra by fixing the inclination angle to the upper limit value of the model (i.e., $\theta_{\rm inc} =$ 87$^{\circ}$ for the {\sc{torus}} model, and $\theta_{\rm inc} =$90$^{\circ}$ for the MY{\sc{torus}} model), to simulate an edge-on inclination torus. For the MY{\sc{torus}} model, we simply modelled the spectra by coupling all of the parameters of {\sc{myts}} and {\sc{mytl}} to {\sc{mytz}}. The relative normalisations of {\sc{myts}} and {\sc{mytl}} with respect to {\sc{mytz}} were fixed to 1. In addition to this torus component, we also added other components to model the spectra at lower energies ($E$ $\lesssim$ 2 keV). These include the ``{\sc{apec}}" component \citep{Smith01} to model the thermal emission from a hot interstellar medium, a power-law component to simulate the scattered emission from the AGN, and the model component for the off-nuclear source, with its power-law normalisation allowed free to vary to account for any flux variations of the source between the \textsl{Chandra} and \textsl{NuSTAR} observations.\footnote{We note that the flux normalisation of the off-nuclear source is consistent with that measure using the \textsl{Chandra} data alone with smaller aperture size.}  


We are able to get a good fit to the data using both models, with the {\sc{torus}} model having a marginally better fit statistic than the MY{\sc{torus}} model ($\chi^{2}$/d.o.f $=$ 368/317 for the {\sc{torus}} model, and $\chi^{2}$/d.o.f $=$ 373/318 for the MY{\sc{torus}} model). In Table 3, we detail the results measured by both models, which agree very well with each other. The models infer a photon index and column density of $\Gamma \approx$ 1.9 and $N_{\rm{H}}$ $\approx$ 2.0 $\times$ 10$^{23}$ cm$^{-2}$. The photon index measured is close to the median value found for the overall (non-blazar) \textsl{Swift}-BAT AGN; i.e., 1.78 $\pm$ 0.01 \citep{Ricci17}. The column density measured indicates that ESO 121-G6 is a heavily obscured AGN, but not CT. The scattering fraction measured with respect to the intrinsic power-law is small; i.e., $f_{\rm scatt} \ll$ 1$\%$, but consistent with that found in many other obscured AGN (e.g., \citealp{Noguchi10}; \citealp{Gandhi14}; \citealp{Gandhi15}). The {\sc{apec}} thermal component indicates a plasma temperature of kT $\approx$ 0.2 keV. Based on these models, we calculated observed and intrinsic luminosities of $L_{\rm 2-10\ keV,\ obs} \approx$  3.4 $\times$ 10$^{40}$ erg s$^{-1}$ and $L_{\rm 2-10\ keV,\ int} \approx$ 1.0 $\times$ 10$^{41}$ erg s$^{-1}$, respectively. The intrinsic luminosity we measured is consistent with that inferred from the AGN mid-IR luminosity (see Section 3), supporting the results from our X-ray spectral analysis. In Figure 3, we show the broadband X-ray spectrum of the AGN and our best-fit models.

The observed luminosity we measured for the AGN is about two orders of magnitude higher than that measured in the archival \textsl{XMM-Newton} data (i.e., $L_{\rm 2-10\ keV,\ obs} = $ 5.37 $\times$ 10$^{38}$ erg s$^{-1}$), suggesting substantial variability between the \textsl{XMM-Newton} observation and our more recent X-ray observations of the source over a period of a decade (2007 and 2017, respectively; see Figure 4). The observed photon index of the AGN has also become significantly harder over this 10-year period (i.e., from $\Gamma_{\rm 2-10\ keV} =$ 1.74$^{+1.14}_{-1.26}$ to $\Gamma_{\rm 2-10\ keV} =$ 0.16 $\pm$ 0.07). These results indicate that the AGN has varied between the 2007 \textsl{XMM-Newton} observation and our 2017 \textsl{Chandra} and \textsl{NuSTAR} observations, both in terms of luminosity and spectral shape, suggesting that ESO 121-G6 could be a candidate for an X-ray changing-look AGN.

\begin{figure}
\centering
\includegraphics[scale=0.425]{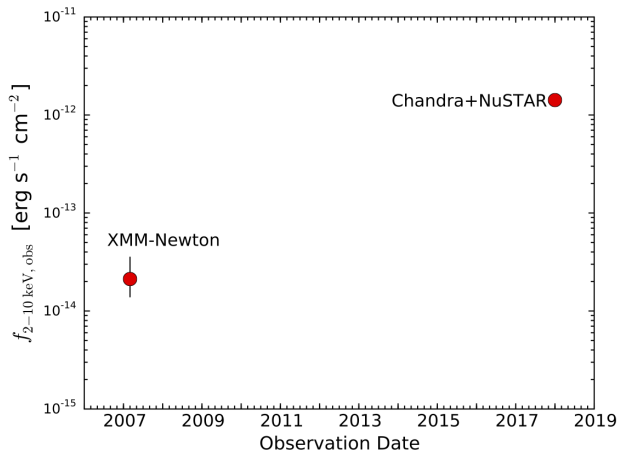}
\caption[[\textsl{Chandra} and \textsl{NuSTAR} images of ESO 121-G6 and the broadband X-ray spectrum]{\footnotesize X-ray flux of the AGN in ESO 121-G6 in the 2--10 keV band observed by \textsl{XMM-Newton} in 2007, and \textsl{Chandra} and \textsl{NuSTAR} in 2017. The significant difference between the two fluxes suggest that the source could be a candidate for a changing-look AGN in the X-rays.}
\end{figure}


\subsection{NGC 660}

\begin{figure*}
\begin{center}
\includegraphics[scale=0.38]{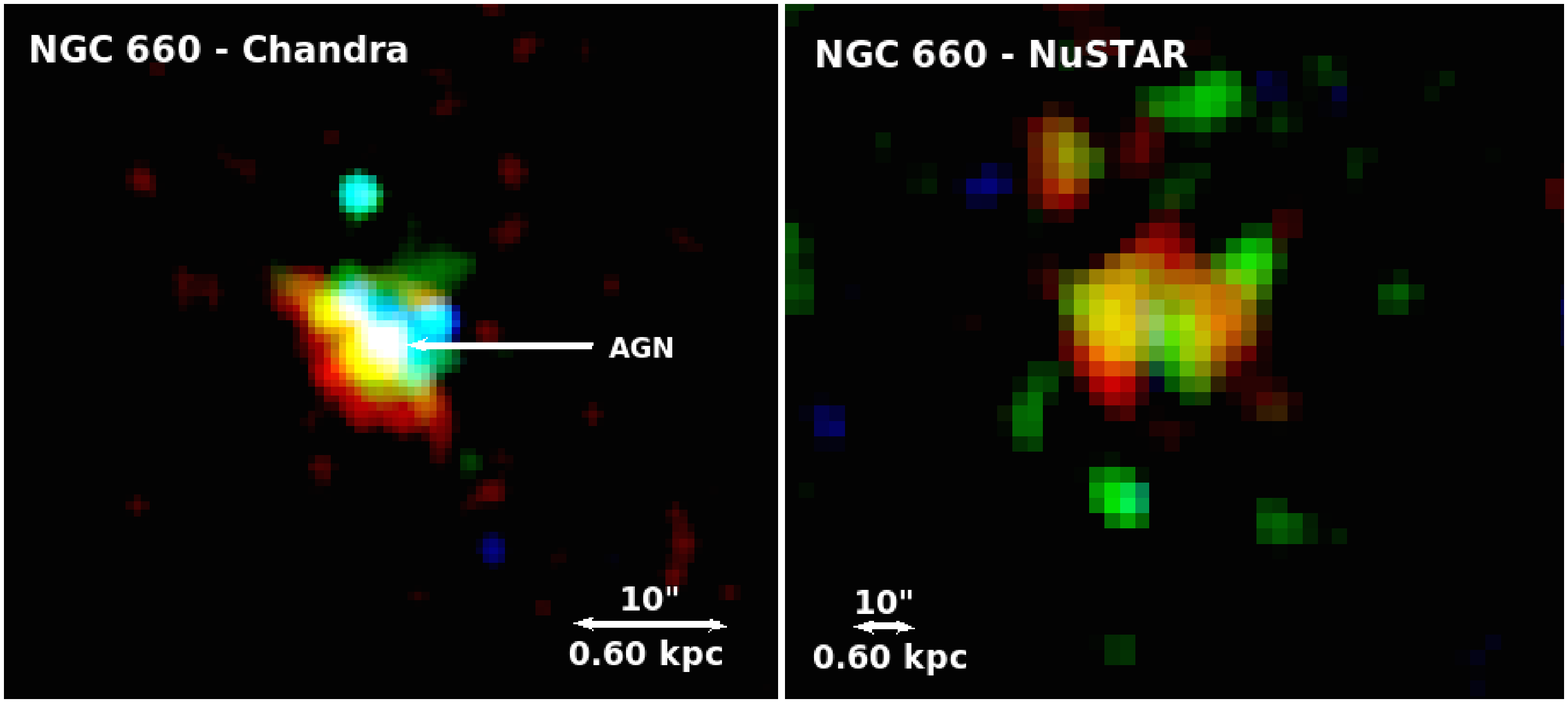} \\ \vspace{+1em}
\includegraphics[scale=0.35]{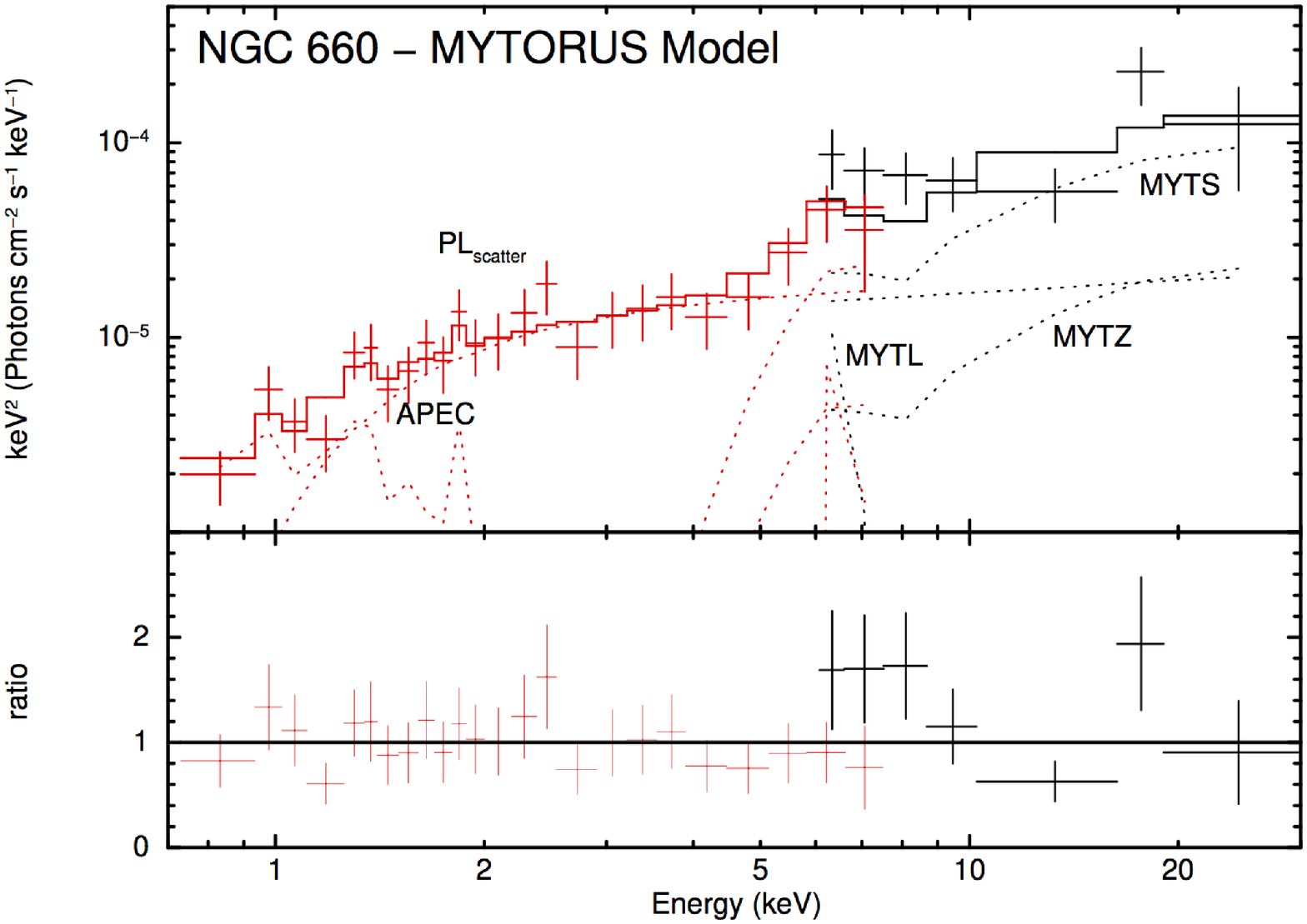}
\includegraphics[scale=0.35]{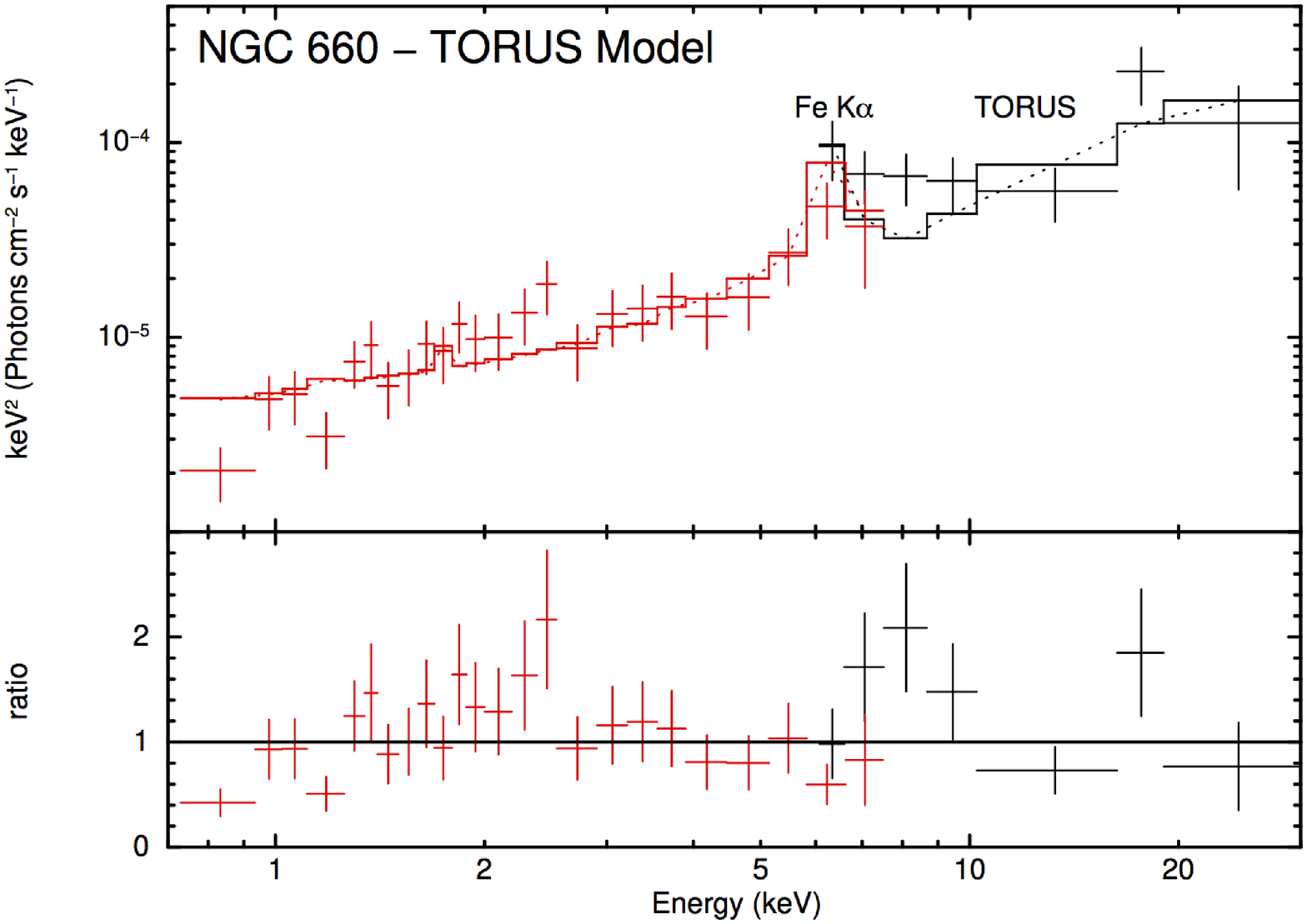}
\caption[\textsl{Chandra} and \textsl{NuSTAR} images of NGC 660 and the broadband X-ray spectrum]{\footnotesize \emph{Top}: \textsl{Chandra} and \textsl{NuSTAR} RGB images of NGC 660 (\textsl{Chandra} - Red: 0.5--1 keV, Green: 1--2 keV, Blue: 2--8 keV; \textsl{NuSTAR} - Red: 3--8 keV, Green: 8--24 keV, Blue: 24--30 keV). The spectra were extracted using a 20$\arcsec$-radius region for \textsl{NuSTAR} and 1.5$\arcsec$-radius for \textsl{Chandra} (to minimise contamination from off-nuclear emission), centred on the radio position of the AGN \citep{Argo15}. The images are smoothed with a Gaussian function of radius 3 pixels, corresponding to 1.5$\arcsec$ and and 7.4$\arcsec$ for \textsl{Chandra} and \textsl{NuSTAR}, respectively. North is up and east is to the left in all images. \emph{Bottom}: Best-fitting MY{\sc{torus}} (left) and {\sc{torus}} (right) models to the combined \textsl{NuSTAR} (black) and \textsl{Chandra} (red) data of NGC 660. The data have been rebinned to a minimum of 3$\sigma$ significance with a maximum of 500 bins for visual clarity. The top panel shows the data and unfolded model in $E^{2}F_{E}$ units, whilst the bottom panel shows the ratio between the data and the folded model. The spectra were fitted using the MY{\sc{torus}} \citep{MY09} and {\sc{torus}} \citep{BN11} models to simulate the obscuring torus, an {\sc{apec}} component as well as a scattered power-law component to model the emission at the softest energies (this component is not visible in the {\sc{torus}} model plot as its contribution is very low). The direct, scattered and line components of the MY{\sc{torus}} model are labelled as {\sc{mytz}}, {\sc{myts}} and {\sc{mytl}}, respectively. The model components fitted to the data are shown as dotted curves, and the combined model fit is shown as solid curves.}
\end{center}
\end{figure*}

NGC 660, located at a distance of 12.3 Mpc, is classified as a rare polar ring galaxy with a LINER-type (low ionization nuclear emission-line region) nuclear spectrum in the optical. In 2013, a high resolution radio observation by e-MERLIN revealed a radio outburst at the centre of the galaxy from a compact and extremely bright continuum source \citep{Argo15}. The radio source, which was not detected in previous radio observations, is probably associated with a newly awoken AGN in the galaxy (\citealp{Argo15}). The galaxy has been observed multiple times in the X-ray band at low energies by \textsl{Chandra} and \textsl{XMM-Newton} between 2001 and 2012 prior to the radio outburst. The \textsl{Chandra} data revealed diffuse emission at the centre of the galaxy, which peaks at the position of the radio source (RA $=$ 01:43:02.32 $\pm$ 1.02 mas, and Dec $=$ 13:38:44.90 $\pm$ 0.78 mas; \citealp{Argo15}). The diffuse X-ray emission heavily contaminates the central source emission up to $\sim$ 6 keV. Although the radio observations indicate that there is significant variability, we found no significant spectral or flux variability between the two sufficiently long \textsl{Chandra} observations ($t_{\rm exp} \geq$ 10 ks), conducted in 2012 November (ObsID 15333; exposure time, $t_{\rm exp} =$ 23.1 ks) and 2012 December (ObsID 15587; exposure time, $t_{\rm exp} =$ 28.1 ks); i.e., $f_{\rm 0.5-8, obs} =$ 5.12$^{+1.65}_{-2.31}$ $\times$ 10$^{-14}$ erg s$^{-1}$ cm$^{-2}$ and $f_{\rm 0.5-8, obs} =$ 6.29$^{+1.92}_{-2.49}$ $\times$ 10$^{-14}$ erg s$^{-1}$ cm$^{-2}$ , respectively. Comparing the X-ray flux of the AGN measured from the archival \textsl{Chandra} data with multiwavelength intrinsic luminosity indicators such as radio continuum and narrow-line region emission lines (measurements taken after and before the outburst, respectively) indicates that the source emission is severely suppressed in X-rays, suggesting heavy obscuration.

\subsubsection{High Spatial Resolution Mid-IR Observation}

NGC 660 was observed at high angular resolution in the mid-IR band using the upgraded Very Large Telescope (VLT) Imager and Spectrometer for mid-IR (VISIR; field of view 38$\arcsec$ $\times$ 38$\arcsec$; 0.045 arcsec pixel$^{-1}$; \citealp{Kaufl15}; \citealp{Kerber16}). The source was observed on 2018-08-15 (Program ID 0101.B-0386(A); PI A. Annuar) for 1 hour (on-source time), using the J8.9 filter ($\lambda =$ 7.8--9.5 $\mu$m) in parallel chop and nod mode. The data were reduced with the custom made python pipeline, VISIC $\&$ Isaac Pipeline Environment (VIPE; Asmus, in prep.; https://github.com/danielasmus/vipe), and flux calibrated using the consecutively observed standard stars HD 22663 and HD 26967 from \cite{Cohen99} standard catalogue. Similar to ESO 121-G6, a compact nucleus is detected, but in NGC 660, the compact emission is surrounded by relatively bright host emission (an almost edge-on starburst ring; see Figure 2). We measured the flux of the unresolved core using a manual point source function scaling at 8.9$\mu$m, which was then converted to 12$\mu$m using a correction factor of 1.33 $\pm$ 0.21, assuming the typical Type 2 Seyfert mid-IR spectral energy distribution from \citet{Asmus14}. The resulting 12$\mu$m flux density measured is 32 $\pm$ 11 mJy, corresponding to a luminosity of $L_{\rm 12\mu m} =$ (1.5 $\pm$ 0.6) $\times$ 10$^{41}$ erg s$^{-1}$. 

\subsubsection{X-ray Observations $\&$ Data Extraction}

We observed NGC 660 at hard X-ray energies with \textsl{NuSTAR} in 2015 (after the radio outburst) for 56.0 ks (2015-08-23; ObsID 60101102002), contemporaneously with a short \textsl{Chandra} observation (ObsID 18352; $t_{\rm exp} =$ 10.1 ks) to further check for potential variability of the nuclear source at X-ray wavelengths. We reduced our new \textsl{Chandra} data of the source, and compared it to the two archival \textsl{Chandra} observations mentioned earlier. Although the count rate measured for our data is lower than that measured in the two archival data sets, they are consistent with each other within the statistical uncertainty. Therefore, we combined the event files for the observations together using {\sc{xselect}} ($t_{\rm exp, tot} =$ 60.5 ks), and produced the resultant \textsl{Chandra} images of the source at 0.5--3 keV and 3--8 keV using {\sc{dmcopy}}. The images are shown in Figure 5. We extracted the total spectrum of the AGN from a small circular region of 1.5$\arcsec$-radius centred on the radio position, to exclude as much contamination as possible from the complex non-AGN emission. A 20$\arcsec$-radius circular aperture was used to extract the background counts from an offset, source-free region. The total count rate measured from the combined \textsl{Chandra} data in the 0.5--8 keV band is 4.05 $\times$ 10$^{-3}$ counts s$^{-1}$. 

The AGN is detected in both of the \textsl{NuSTAR} FPMs. We show the combined RGB image of the AGN from the two FPMs in Figure 5. The \textsl{NuSTAR} spectrum of NGC 660 from each FPM was extracted using a circular aperture of 20$\arcsec$-radius (corresponding to $\sim$30$\%$ of the \textsl{NuSTAR} encircled energy fraction) centred on the radio position of the AGN. The aperture size was chosen to minimise contamination from off-nuclear sources observed in the \textsl{Chandra} data. The background photons were collected from an annular segment centred on the AGN with inner and outer radii of 50$\arcsec$ and 100$\arcsec$, respectively. We detected significant counts up to $\sim$30 keV from the combined FPM spectrum, and measured a net count rate of 1.41 $\times$ 10$^{-3}$ counts s$^{-1}$ in the 3--30 keV band. However, the \textsl{Chandra} data revealed that our \textsl{NuSTAR} spectrum is heavily contaminated by off-nuclear emission up to $E\sim$6 keV. We therefore excluded the \textsl{NuSTAR} data below this energy for our analysis. The \textsl{NuSTAR} count rate of the AGN in the 6--30 keV band is 8.83 $\times$ 10$^{-4}$ counts s$^{-1}$. 

\subsubsection{X-ray Spectral Fitting}

As mentioned in Section 2.2.2, for our X-ray spectral analysis of NGC 660, we only modelled the \textsl{NuSTAR} spectrum above 6 keV as we expect the data to be heavily contaminated by non-AGN emission below this energy, as indicated by the \textsl{Chandra} data. However for \textsl{Chandra}, we modelled the full energy range (0.5--8 keV) as the spectrum was extracted from a much smaller region, significantly reducing contamination from off-nuclear emission. We first fitted the \textsl{Chandra} and \textsl{NuSTAR} spectra of NGC 660 simultaneously between 3--30 keV using a simple power-law model, absorbed by the Galactic column ($N_{\rm{H}}^{\rm{Gal}}$ $=$ 4.64 $\times$ 10$^{20}$ cm$^{-2}$; \citealp{Kalberla05}). The redshift was fixed at $z =$ 0.003896 in the spectral analysis. The best fitting photon index measured from the spectra is relatively flat; i.e., $\Gamma$ $=$ 0.69 $\pm$ 0.19 (C-stat/d.o.f. $=$ 206/236), indicating significant X-ray absorption along our l.o.s. An excess of emission at $E \sim$ 6 keV also suggests the presence of Fe K$\alpha$ line emission. Adding a Gaussian component to our fit to model the emission indeed confirm a significant Fe K$\alpha$ emission centred at $E =$ 6.49$^{+0.10}_{-0.28}$ keV, with an equivalent width of $EW_{\rm Fe\ K\alpha} =$ 0.65$^{+0.58}_{-0.60}$ keV, indicating significant obscuration, possibly CT if we account for the upper limit uncertainty of the line equivalent width ($EW_{\rm Fe\ K\alpha} \geq$ 1 keV for CT; e.g., \citealp{Maiolino98}; \citealp{Comastri04}).

We then proceeded to fit the broadband X-ray spectrum of NGC 660 (0.5--30 keV) with the MY{\sc{torus}} and {\sc{torus}} models, similar to ESO 121-G6 (see Section 2.1). For the MY{\sc{torus}} model, we also include a scattered power-law component to model the low energy part of the AGN spectrum, and the {\sc{apec}} component to model the non-AGN component at low energies. However for the {\sc{torus}} model, we only included the scattered power-law component as the {\sc{apec}} component could not be constrained, and we are able to obtain a good fit without this component. We fixed the photon index in both models to $\Gamma$ $=$ 1.8 (approximately the typical mean value for the \textsl{Swift}-BAT AGN; \citealp{Ricci17}) as it could not be constrained. 

For the MY{\sc{torus}} model, we measured a column density of $N_{\rm{H}}$ $=$ 6.09$^{+3.19}_{-2.43}$ $\times$ 10$^{23}$ cm$^{-2}$ (C-stat/d.o.f $=$ 182/186), indicating a heavily obscured AGN, close to the CT regime within the statistical uncertainties. The scattering fraction measured for the AGN is relatively high; i.e., $f_{\rm scatt}$ 14$^{+11}_{-6} \%$; however we note that, given the relatively low quality data, this scattered component will also include contributions from other processes such as unresolved X-ray binaries, meaning that the intrinsic scattering fraction could be smaller than this value. The plasma temperature measured by the {\sc{apec}} model component is kT $=$ 0.34$^{+0.79}_{-0.16}$ keV. The observed and intrinsic luminosities measured from this model are $L_{\rm 2-10\ keV,\ obs} =$ 1.19 $\times$ 10$^{39}$ erg s$^{-1}$ and $L_{\rm 2-10\ keV,\ int} =$ 5.74 $\times$ 10$^{39}$ erg s$^{-1}$, respectively. The intrinsic luminosity measured using this model is significantly lower than that predicted by the X-ray:12$\mu$m intrinsic relationship derived by \citet{Asmus15} (see Section 3 $\&$ Figure 9), suggesting that the obscuring column may be higher than what we measured from this model, and is potentially CT.

We are able to also get an equally good fit with the {\sc{torus}} model (C-stat/d.o.f $=$ 206/187). However, the results obtained are quite different than the MY{\sc{torus}} model. The column density measured by this model is in the CT regime; i.e., $N_{\rm{H}}$ $\geq$ 5.38 $\times$ 10$^{24}$ cm$^{-2}$, and the intrinsic luminosity inferred after accounting for this extreme obscuration is $L_{\rm 2-10\ keV,\ int} \geq$ 1.74 $\times$ 10$^{41}$ erg s$^{-1}$. This luminosity is consistent with that predicted by the X-ray:12$\mu$m correlation (see Section 3 $\&$ Figure 9). In Figure 5, we show the broadband X-ray spectrum of the AGN in NGC 660 and our best fit models.

\begin{figure*}
\begin{center}
\includegraphics[scale=0.38]{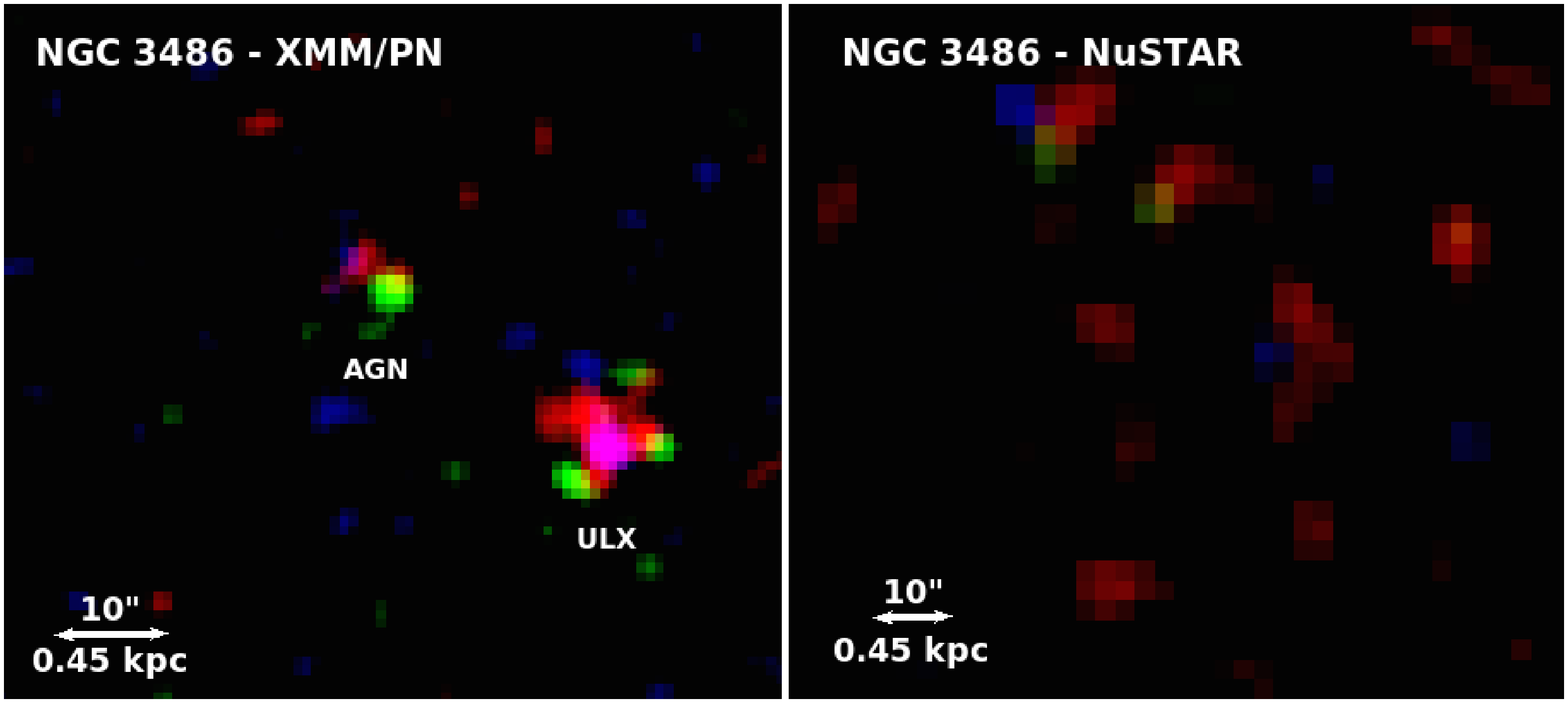} 
\includegraphics[scale=0.375]{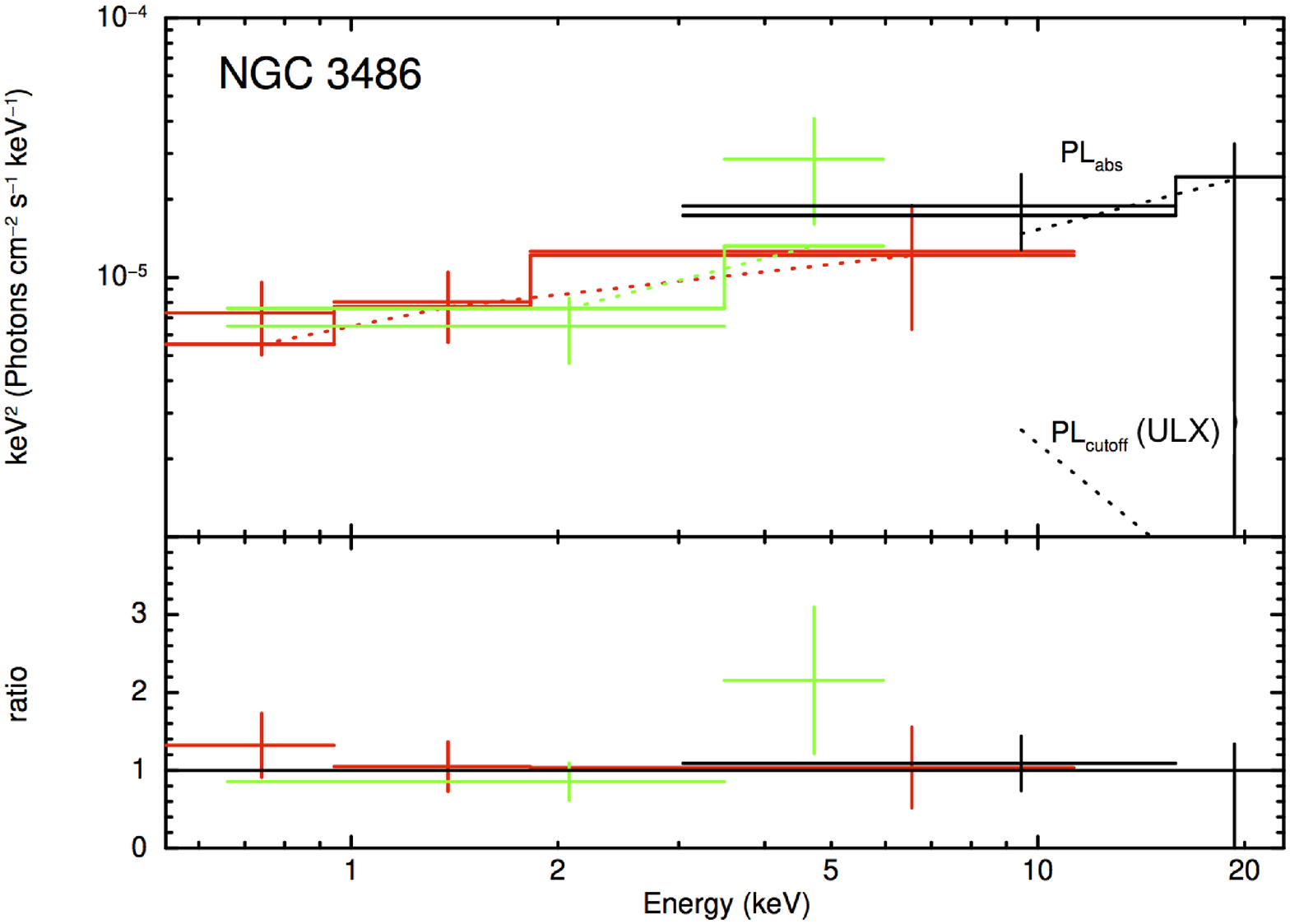}
\caption[\textsl{XMM-Newton} and \textsl{NuSTAR} images of NGC 3486 and the broadband X-ray spectrum]{\small \emph{\emph{Top}}: \textsl{XMM-Newton} and \textsl{NuSTAR} RGB images of NGC 3486 (\textsl{XMM-Newton} - Red: 0.5--1 keV, Green: 1--2 keV, Blue: 2--10 keV; \textsl{NuSTAR} - Red: 3--10 keV, Green: 10--15 keV, Blue: 15--24 keV). The spectra were extracted using a 20$\arcsec$-radius region for both \textsl{NuSTAR} and \textsl{XMM-Newton}, centred on the optical position of the AGN. The images are smoothed with a Gaussian function of radius 3 pixels, corresponding to 9.6$\arcsec$ and and 7.4$\arcsec$ for \textsl{Chandra} and \textsl{NuSTAR}, respectively. North is up and east is to the left in all images. \emph{\emph{Bottom}}: Best-fitting model to the combined \textsl{NuSTAR} (black), \textsl{XMM-Newton} PN (red) and MOS1+2 (green) data of NGC 3486. The data have been rebinned to a minimum of 3$\sigma$ significance with a maximum of 500 bins for visual clarity. The spectra were fitted using an absorbed power-law model to simulate the AGN emission, and a cut-off power-law to account for possible contamination by the ULX in the \textsl{NuSTAR} spectrum. The model components fitted to the data are shown as dotted curves, and the combined model fit is shown as solid curves.}
\end{center}
\end{figure*}

\subsection{NGC 3486}

NGC 3486 is a face-on galaxy located at a distance of 9.2 Mpc and has an optical nuclear spectrum which identifies it as a Type 2 Seyfert \citep{Ho97}. The object had previously been observed in X-rays by \textsl{Chandra} (ObsID 393; 1999-11-03; $t_{\rm exp} =$ 1.8 ks) and \textsl{XMM-Newton} (ObsID 0112550101; 2001-05-09; $t_{\rm exp} =$ 15.2 ks). Whilst a source associated with the optical nuclear position of the galaxy is detected in the \textsl{XMM-Newton} data, along with an ultraluminous X-ray source (ULX) $\sim$23$\arcsec$ from the nucleus \citep{Foschini02}, nothing was detected in the short \textsl{Chandra} observations \citep{Ho01}. NGC 3486 is one of four galaxies in \citet{Goulding09} which lacks high-resolution \emph{Spitzer}-IRS spectroscopic data, and therefore no [Ne{\sc{v}}] flux measurement was made. However, an [O{\sc iv}] emission line is clearly detected in the low-resolution \emph{Spitzer}-IRS spectroscopy and the optical spectrum indicates that NGC 3486 hosts an AGN. Previous works have performed X-ray spectral analysis on the \textsl{XMM-Newton} data of NGC 3486 and found some evidence for it to be heavily obscured; e.g., a relatively flat power-law photon index at 2--10 keV (\citealp{Cappi06}; \citealp{Brightman08}).

\subsubsection{Mid-IR Observation}

The central part of NGC 3486 has been observed at high spatial resolution in the mid-IR ($N$-band filters) in 2010 with the Michelle instrument, mounted on the Gemini-North telescope (0.1005 arcsec pixel$^{-1}$). However, nothing was detected in the images \citep{Asmus14}. The upper limit on the 12$\mu$m flux density derived from the data is 5.1 mJy. This corresponds to an upper limit luminosity of $L_{\rm 12\mu m}$ $<$ 1.3 $\times$ 10$^{40}$ erg s$^{-1}$.

\subsubsection{X-ray Observations $\&$ Data Extraction}

We observed NGC 3486 at hard X-ray energies with \textsl{NuSTAR} on 2015-01-26 for $t_{\rm exp} =$ 28.9 ks (ObsID 60001150002). The observation was coordinated with a short \textsl{Swift} X-ray Telescope (XRT; \citealp{Burrows05}) observation to facilitate our X-ray analysis at low energies (ObsID 00080813001; $t_{\rm exp} \sim$ 5 ks). Neither the AGN nor the ULX in NGC 3486 were significantly detected in either our \textsl{NuSTAR} or \textsl{Swift}-XRT observations using the detection technique adopted in other \textsl{NuSTAR} studies of faint sources (significance $\lesssim$ 2.6$\sigma$; e.g., \citealp{Luo13}, \citealp{Lansbury14}; \citealp{Stern14}).{\footnote{The XRT data were reduced using the {\sc{xrtpipeline}} v0.13.0, which is part of the XRT Data Analysis Software ({\sc{xrt-das}}) within {\sc{heasoft}}.}} The upper limit to the count rates measured for the AGN in the XRT and \textsl{NuSTAR} observations (total for both FPMs) are CR$_{\rm 0.5-10, AGN} <$ 3.75 $\times$ 10$^{-3}$ counts s$^{-1}$ and CR$_{\rm 3-24, AGN} <$ 0.74 $\times$ 10$^{-3}$counts s$^{-1}$ at 0.5--10 keV and 3--24 keV, respectively. Based on {\sc{webpimms}}{\footnote{{\sc{webpimms}} is a mission count rate simulator tool, available online at https://heasarc.gsfc.nasa.gov/cgi-bin/Tools/w3pimms/w3pimms.pl .}}, the XRT flux corresponds to an upper limit flux of $f_{\rm 0.5-10, AGN} <$ 1.49 $\times$ 10$^{-13}$ erg s$^{-1}$ cm$^{-2}$, assuming $z = $ 0.00272, $N_{\rm{H}}^{\rm{Gal}}$ $=$ 1.90 $\times$ 10$^{20}$ cm$^{-2}$ \citep{Kalberla05}, and $\Gamma = $ 1.8. The flux measured from the \textsl{XMM-Newton} data; i.e., $\sim$ 1.3 $\times$ 10$^{-13}$ erg s$^{-1}$ cm$^{-2}$ \citep{Cappi06}, is consistent with this upper limit, which could be an indication that the AGN has not significantly varied; however, we could not rule out variability toward lower fluxes. The upper limit count rate measured for the ULX is CR$_{\rm 0.5-10, ULX} <$ 4.41 $\times$ 10$^{-3}$ counts s$^{-1}$, corresponding to an upper limit flux of $f_{\rm 0.5-10, ULX} <$ 1.66 $\times$ 10$^{-13}$ erg s$^{-1}$ cm$^{-2}$. The flux measured for the ULX from the archival \textsl{XMM-Newton} data; i.e., $\sim$ 7.7 $\times$ 10$^{-14}$ erg s$^{-1}$ cm$^{-2}$ \citep{Foschini02}, is also consistent with the upper limit measured with our new data.

Although the AGN in NGC 3486 is not significantly detected in our \textsl{NuSTAR} data, we extracted the spectra to assist our broadband X-ray spectral analysis of the source with the archival \textsl{XMM-Newton} data. We extracted the \textsl{XMM-Newton} spectra of the AGN from the three EPIC cameras (PN, MOS1 and MOS2) using a circular source region of 10$\arcsec$ in radius (to avoid the ULX), centred on the optical position of the AGN obtained from the Sloan Digital Sky Survey Catalog (RA $=$ 11:00:23.87, Dec $= +$28:58:30.49). The background photons were measured in an annular segment around the source, with inner and outer radii of 15$\arcsec$ and 30$\arcsec$, respectively, avoiding the ULX emission. For the \textsl{NuSTAR} data, we extracted the spectra from both FPMs using a 20$\arcsec$-radius circular region for the source, and an annular segment with inner and outer radii of 50$\arcsec$ and 100$\arcsec$ for the background spectrum. For our spectral analysis, we did not use the \textsl{Swift}-XRT data as they do not provide additional constraints beyond those already achieved by the \textsl{XMM-Newton} and \textsl{NuSTAR} data. We show the X-ray RGB images of NGC 3486 in Figure 6. 

\subsubsection{X-ray Spectral Fitting}

Due to the low quality of the archival \textsl{XMM-Newton} data and our \textsl{NuSTAR} data of NGC 3486 ($\sim$75 counts and not significantly detected, respectively), we modelled its spectra between 0.5--24 keV using a simple power-law model, absorbed by a fixed Galactic column, $N_{\rm{H}}^{\rm{Gal}}$ $=$ 1.90 $\times$ 10$^{20}$ cm$^{-2}$ \citep{Kalberla05}, with an additional absorption component, {\sc{zwabs}}, to simulate the intrinsic absorption of the source. We fixed the redshift to $z =$ 0.002272 in our model. Although the nearby ULX is not significantly detected in the \textsl{NuSTAR} data, faint emission consistent with the ULX position can be seen in the \textsl{NuSTAR} 3--8 keV image (see Figure 6). Therefore, we include a cut-off power-law component in our model to account for the ULX contribution in the \textsl{NuSTAR} data. We set the photon index of the ULX to that measured by \cite{Foschini02} using the \textsl{XMM-Newton} data (i.e., $\Gamma_{\rm ULX} =$ 2.2), assuming that it has not significantly varied between the \textsl{XMM-Newton} and \textsl{NuSTAR} observations, and fixed the X-ray energy cut-off at 10 keV, consistent with that found in other ULXs (e.g., \citealp{Walton13}; \citealp{Bachetti13}; \citealp{Walton14}; \citealp{Rana15}). The flux normalisation of the component was left free to vary.\footnote{We note that the flux normalisation of the off-nuclear source is slightly lower than that measured in the \textsl{Chandra} data alone using smaller aperture, but is consistent within 99$\%$ confidence level.}  We did not include any additional components (e.g., {\sc{apec}}) to model potential additional non-AGN emission at the soft energies as they could not be constrained. 

The best fitting photon index and column density measured by our model for the AGN are $\Gamma$ $=$ 1.52$^{+0.43}_{-0.24}$ and $N_{\rm{H}}$ $\leq$ 1.37 $\times$ 10$^{21}$ cm$^{-2}$ (C-stat/d.o.f $=$ 27/34), respectively. The column density measured indicates that NGC 3486 is unobscured, and is a very low luminosity AGN with an intrinsic luminosity of $L_{\rm 2-10\ keV,\ int} =$ 3.84 $\times$ 10$^{38}$ erg s$^{-1}$. If we fitted our model only to the \textsl{XMM-Newton} data, in which the AGN and the ULX are better resolved, we obtained consistent results. Our analysis disagree with the previous suggestions that the source is likely to be CT by previous low X-ray energy studies (\citealp{Cappi06}; \citealp{Brightman08}). Furthermore, we predict that the AGN would have been about an order of magnitude brighter in \textsl{NuSTAR} 3-24 keV band if it were CT (assuming the {\sc{torus}} model with $N_{\rm{H}}$ $=$ 1.5 $\times$ 10$^{24}$ cm$^{-2}$). However, the non-detection of the source in the \textsl{NuSTAR} observation could also indicate a heavily Compton-thick source with $N_{\rm{H}}$ $\geq$ 10$^{25}$ cm$^{-2}$.

We show the spectra for NGC 3486 and our best-fit model in Figure 6. The unobscured nature of NGC 3486 as revealed by our spectral analysis contradicts with the optical type 2 classification of the source. Therefore, the AGN could be a candidate for a ``true'' type 2 AGN; i.e., an AGN in which the BLR is genuinely absent (e.g., \citealp{Panessa09}; \citealp{Trump11}; \citealp{Bianchi12}). However, it should be noted that high sensitivity observations have recently revealed BLRs in sources previously thought to be true type 2 AGN (e.g., \citealp{Bianchi19}).

\subsection{NGC 5195}

NGC 5195 (also known as M51b), located at a distance of 8.3 Mpc, is an irregular galaxy interacting with NGC 5194 (M51a). The nucleus is classified as a LINER in the optical band, and the galaxy has been observed numerous times in X-rays by \textsl{Chandra}, \textsl{XMM-Newton} and \textsl{NuSTAR}, most of the time as part of the overall M51 system. Whilst much of the attention of previous studies have focused on its brighter and more picturesque companion, NGC 5195 came to widespread attention in recent years when it was observed to undergo a violent X-ray outburst, potentially associated with AGN feedback \citep{Schlegel16}. Its companion, NGC 5194, has been confirmed to be a low luminosity CT AGN using multiple X-ray data sets including \textsl{NuSTAR} \citep{Xu16}. A variable ULX (ULX-7), identified in the northern spiral arm of NGC 5194, was recently identified to be powered by neutron star (\citealp{Rodriguez19}; \citealp{Brightman19}). We observed NGC 5195 as part of the M51 system with \textsl{NuSTAR} in 2017 \citep{Brightman18}.  Based on Figure 1, the observed X-ray luminosity of the source measured from archival data is significantly lower than that expected from the [Ne{\sc{v}}] line emission, suggesting that the nuclear source is heavily buried. 

\begin{figure*}
\begin{center}
\includegraphics[scale=0.38]{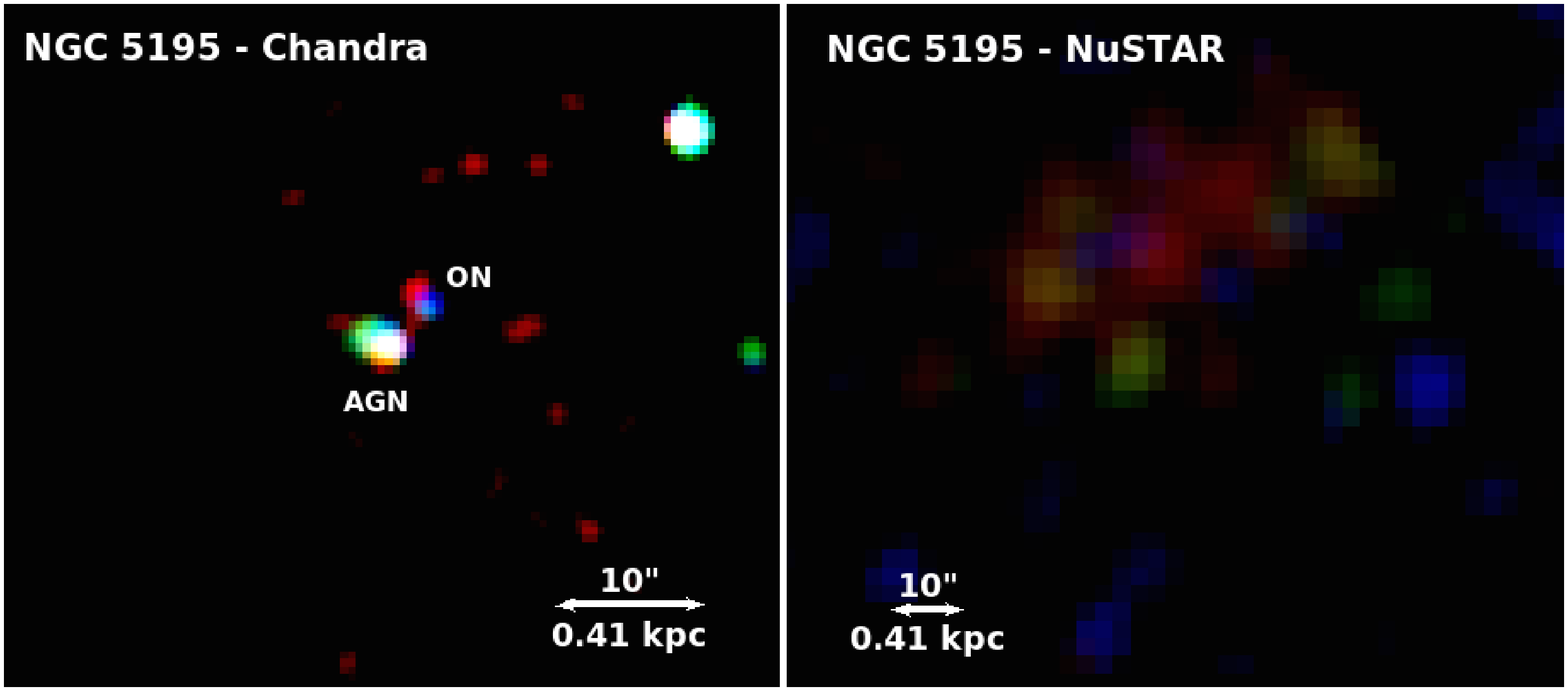}
\includegraphics[scale=0.375]{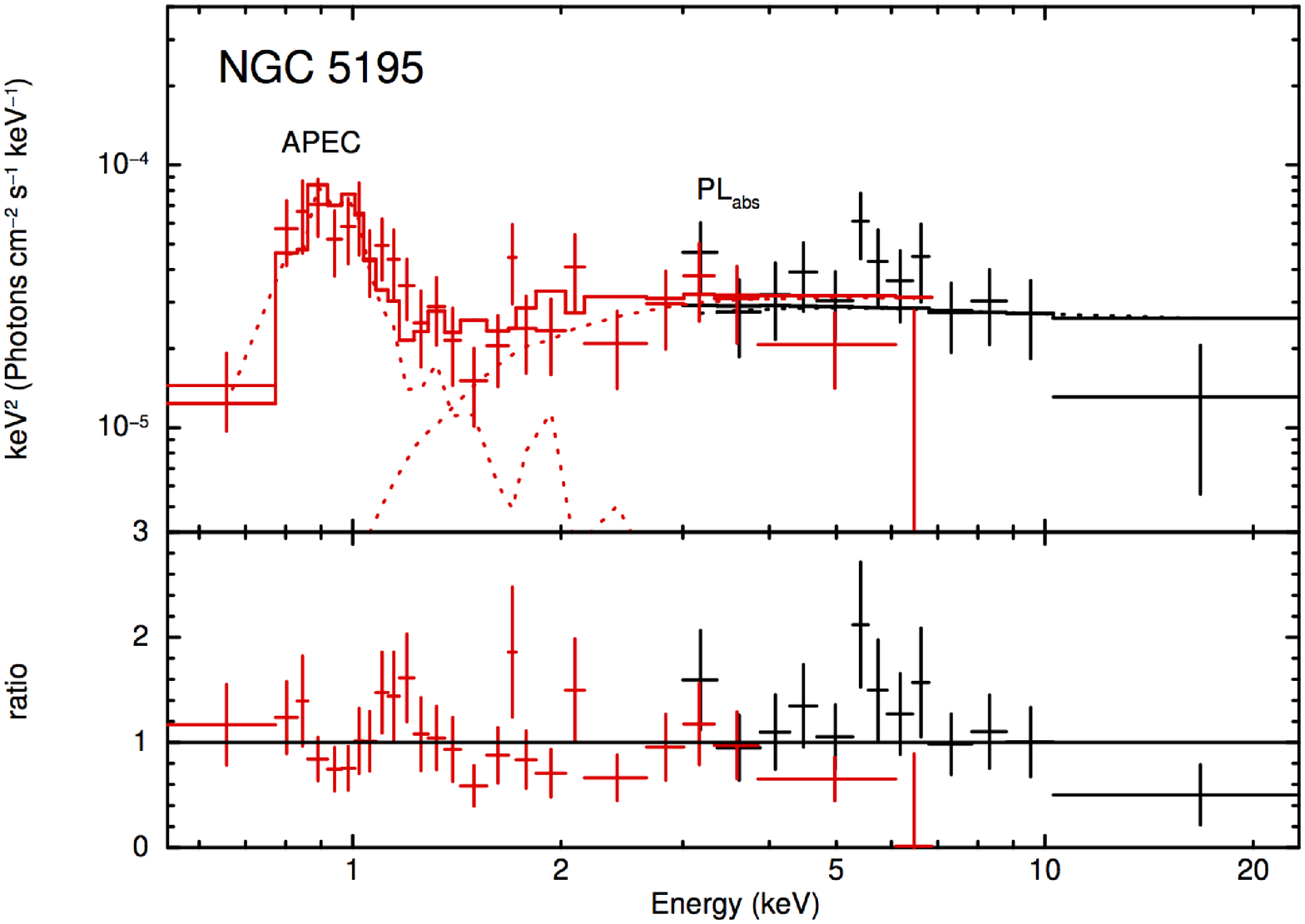}
\caption[\textsl{Chandra} and \textsl{NuSTAR} images of NGC 5195 and the broadband X-ray spectrum]{\small \emph{\emph{Top}}: \textsl{Chandra} and \textsl{NuSTAR} RGB images of NGC 5195 (\textsl{Chandra} - Red: 0.5--1 keV, Green: 1--2 keV, Blue: 2--8 keV; \textsl{NuSTAR} - Red: 3--10 keV, Green: 10--15 keV, Blue: 15--24 keV). The spectra were extracted using a 20$\arcsec$-radius region for both \textsl{NuSTAR} and \textsl{Chandra}. The images are smoothed with a Gaussian function of radius 3 pixels, corresponding to 1.5$\arcsec$ and and 7.4$\arcsec$ for \textsl{Chandra} and \textsl{NuSTAR}, respectively. North is up and east is to the left in all images. \emph{Bottom}: Best-fitting model to the combined \textsl{NuSTAR} (black) and \textsl{Chandra} (red) data of NGC 5195. The data have been rebinned to a minimum of 3$\sigma$ significance with a maximum of 500 bins for visual clarity. The spectra were fitted using an absorbed power-law model, an {\sc{apec}} component to model the emission at the softest energies, and an additional power-law component to model the off-nuclear source at the north-west of the AGN (labelled as ON in the image - this component is not visible in the plot as the flux normalisation is very low). The model components fitted to the data are shown as dotted curves, and the combined model fit is shown as solid curves.}
\end{center}
\end{figure*}


\subsubsection{High Spatial Resolution Mid-IR Observation}

As opposed to its more famous companion NGC 5194, NGC 5195 has not been observed at mid-IR wavelengths at high spatial resolution. However it is detected by {\it{WISE}} \citep{Wright10}, and the W3 band (12$\mu$m) luminosity measured from the profile fitting photometry, tracing the unresolved component, is $L_{\rm 12\mu m, WISE}$ $=$ 7.94 $\times$ 10$^{41}$ erg s$^{-1}$ \citep{Cutri13}, which we regard here as upper limit on the nuclear mid-IR flux since the low angular resolution of {\it{WISE}} means that the host galaxy could contaminate the mid-IR emission from the AGN.

\subsubsection{X-ray Observations $\&$ Data Extraction}

We observed the M51 system with \textsl{NuSTAR} (PI M. Brightman; ObsID 60201062002 \& 60201062003; $t_{\rm exp} =$ 47.2 \& 163 ks, respectively), coordinated with \textsl{Chandra} (ObsID 19522; 38.2 ks) in March 2017. Our \textsl{Chandra} data revealed two point sources within $\sim$5$\arcsec$ of the radio position of NGC 5195 (RA $=$ 13:29:59.534 $\pm$ 1 mas, and Dec. $=$ $+$47:15:57.33 $\pm$ 10 mas; \citealp{Rampadarath18}), which were not resolved in previous observations (e.g., \citealp{Terashima01}; \citealp{Schlegel16}; see Figure 7). This is likely due to the fact that NGC 5195 has generally been located off-axis in previous \textsl{Chandra} observations, where the point spread function is larger. We used {\sc{wavdetect}} to determine the centroid position of the two central sources in the \textsl{Chandra} energy band of 2--8 keV.  Based on this, we found that one of the \textsl{Chandra} sources is located at position of RA $=$ 13:29:59.41, and Dec. $=$ $+$47:15:57.29, with formal wavdetect errors of 0$\farcs$10 and 0$\farcs$10 (see footnote 7), respectively, consistent with the e-MERLIN position of the nucleus within $\sim$1.5$\arcsec$. Therefore, we used the position of this \textsl{Chandra} source when extracting the X-ray spectra of the AGN. The other point source (off-nuclear) was detected at RA $=$ 13:29:59.21, and Dec. $=$ $+$47:16:00.01, with formal wavdetect errors of 0$\farcs$19 and 0$\farcs$11 (see footnote 7), respectively.

A source near the \textsl{Chandra} position of the AGN was detected in both of the \textsl{NuSTAR} observations, with significant counts up to $\sim$10 keV. The combined \textsl{NuSTAR} RGB image of the source is shown in Figure 7. The count rate detected in the combined \textsl{NuSTAR} data in the 3--10 keV band is 4.59 $\times$ 10$^{-4}$ counts s$^{-1}$. This source could be associated with either the AGN or the off-nuclear source. We extracted the \textsl{NuSTAR} spectra using a 20$\arcsec$-radius circular region adopting the \textsl{Chandra} position of the AGN as the centroid position. The background was extracted using a 100$\arcsec$-radius aperture from an offset source-free region. The \textsl{Chandra} spectrum was also extracted using a 20$\arcsec$-radius circular region to match the \textsl{NuSTAR} extraction region. The background for the \textsl{Chandra} spectrum was extracted using a circular region of 50$\arcsec$-radius from a source-free area. In addition, we also extracted the spectrum of the off-nuclear point source detected within the extraction region of the \textsl{Chandra} data to account for its contribution to the extracted spectra. The source flux was extracted using a small 1.5$\arcsec$-radius region, and the background flux was extracted using a 50$\arcsec$-radius aperture from a source-free region. 

\subsubsection{X-ray Spectral Fitting}

For the spectral analysis of NGC 5195, we did not include its archival \textsl{NuSTAR} data (2012-10-29; ObsID 60002038002; $t_{\rm exp} =$ 16.7 ks) as we found that the source fluxes in the 3--8 keV bands are different from each other within the statistical uncertainties at 90$\%$ confidence limit; i.e., $L_{\rm 3-8\ keV,\ obs} =$ 0.55$^{+0.09}_{-0.07}$ $\times$ 10$^{-13}$ erg s$^{-1}$ cm$^{-2}$ and $L_{\rm 3-8\ keV,\ obs} =$ 1.08$^{+0.49}_{-0.41}$ $\times$ 10$^{-13}$ erg s$^{-1}$ cm$^{-2}$, for our new data and the archival data, respectively. Investigation of the older \textsl{Chandra} data of NGC 5195, in which the nucleus and the off-nuclear source were unresolved by \citet{Rampadarath18}, also revealed that the nucleus has varied by a factor of $\sim$3 between data taken in 2000 and 2012. The flux variability may be attributed to either the AGN or the nearby off-nuclear source. Because of this, we also did not include the older \textsl{Chandra} data in our analysis. 

We first constrained the contribution from the off-nuclear source detected near the AGN in the \textsl{Chandra} data (see Figure 7). We took the same basic approach as for ESO 121-G6 by modelling the source between 0.5--8 keV using a simple power-law model, absorbed by the host galaxy and the Galactic column, $N_{\rm{H}}^{\rm{Gal}}$ $=$ 1.79 $\times$ 10$^{20}$ cm$^{-2}$ \citep{Kalberla05}, assuming that it is located within NGC 5195 ($z =$ 0.001551). The best-fitting photon index and host galaxy absorption measured by the model for the off-nuclear source are $\Gamma$ $=$ 0.48$^{+0.81}_{-0.65}$ and $N_{\rm{H}}$ $<$ 5.61 $\times$ 10$^{21}$ cm$^{-2}$, respectively (C-stat/d.o.f $=$ 26/25). The 0.5--8 keV intrinsic luminosity inferred from the model is $L_{\rm 0.5-8\ keV,\ int} =$ 1.61$^{+0.93}_{-0.33}$ $\times$ 10$^{38}$ erg s$^{-1}$, lower than the luminosity limit for a ULX, suggesting that the off-nuclear source is more likely to be an X-ray binary. The photon index measured, however, is quite low for an X-ray binary (typically $\Gamma \sim$ 2; e.g., \citealp{Yang15}), even after accounting for the statistical uncertainties. 


We then modelled the broadband X-ray spectrum of the AGN between 0.5--24 keV using a power-law model absorbed by the Galactic and intrinsic absorption ({\sc{zwabs}}), an {\sc{apec}} component to model the soft energy emission, and the off-nuclear source component to account for its contribution in the extracted spectrum of the AGN. The photon index and host galaxy absorption of the off-nuclear source were fixed to the values measured earlier, but the flux normalisation parameter was left free to vary in the spectral analysis. Based on this model, we measured an obscuring column density of $N_{\rm{H}}$ $=$ 1.17$^{+1.75}_{-0.99}$ $\times$ 10$^{22}$ cm$^{-2}$ (C-stat/d.o.f $=$ 226/243), indicating that the AGN is just mildly obscured (potentially unobscured within the statistical uncertainties), and is therefore a very low luminosity AGN ($L_{\rm 2-10\ keV,\ int} =$ 6.60 $\times$ 10$^{38}$ erg s$^{-1}$), similar to NGC 3486. The best-fitting photon index measured is $\Gamma$ $=$ 2.12$^{+0.61}_{-0.23}$, and the {\sc{apec}} thermal component indicated a plasma temperature of kT $=$ 0.95$^{+0.08}_{-0.16}$ keV. In general, our results are in agreement with those found by \citet{Brightman18} and \citet{Rampadarath18}, who fitted the same data using slightly different models. We show the broadband X-ray spectra of NGC 5195 in Figure 7. 

\begin{table*}
\caption{X-ray analysis results.}
\centering
\begin{adjustbox}{width=\textwidth,totalheight=\textheight,keepaspectratio}
\begin{tabular}{lcccccccccc}
  \hline
   Name & Facility & Model & Energy Band & $\Gamma$ & $\log N_{\rm H}$ & $\log{L_{\rm 2-10\ keV,\ obs}}$ &  $\log{L_{\rm 2-10 keV, int}}$  & $\chi^{2}$ or C-stat / d.o.f & SC \\
             &   &   & [keV] & & [cm$^{-2}$] & [erg s$^{-1}$] & [erg s$^{-1}$] & &  \\
   (1) & (2)  & (3) & (4) & (5) & (6) & (7) & (8) & (9) & (10)  \\
   \hline
   ESO 121-G6 & \it{C} + \it{N}   & MY{\sc{torus}}  & 0.5--50 & 1.89$\pm$0.08  & 23.33$\pm$0.03  & 40.53 & 41.01 & 373 / 318 ($\chi^{2}$) & 0.10 $\pm$ 0.02  \\ 
                     & \it{C} + \it{N}   & {\sc{torus}} & 0.5--50  & 1.89$^{+0.11}_{-0.06}$  & 23.29 $\pm$ 0.02  & 40.53 & 41.01  & 368 / 317 ($\chi^{2}$) & 0.10 $\pm$ 0.02    \\ 
   NGC 660 & \it{C} + \it{N}   & MY{\sc{torus}}  & 0.5--30 & 1.8$^{f}$  & 23.78$^{+0.18}_{-0.22}$  & 39.07 & 39.76 & 181 / 186 (C-stat) & 0.31 $\pm$ 0.13  \\ 
                    & \it{C} + \it{N} &  {\sc{torus}}    & 0.5--30  & 1.8$^{f}$  & $\geq$ 24.73  & 39.05 & $\geq$ 41.24 & 206 / 187 (C-stat) &0.31 $\pm$ 0.13  \\ 
   NGC 3486 & \it{XMM} + \it{N}  & {\sc{zwabs}}({\sc{zpow}})  & 0.5--24  & 1.52$^{+0.43}_{-0.24}$   & $\leq$21.14 & 38.58  & 38.58 & 27 / 34 (C-stat) & -   \\
   NGC 5195 &  \it{C} + \it{N}   & {\sc{zwabs}}({\sc{zpow}})  & 0.5--24 & 2.12$^{+0.61}_{-0.23}$  & 22.07$^{+0.40}_{-0.81}$  &  38.80 & 38.82 &  226 / 243 (C-stat) & - \\ 
  \hline
\end{tabular}
\end{adjustbox}
  \begin{tablenotes}
   \footnotesize
   \item \emph{Notes.} Column (1) AGN name; (2) X-ray facilities used in the analysis (C: \textsl{Chandra}; N: \textsl{NuSTAR}; XMM: \textsl{XMM-Newton}); (3) best-fit models to the spectra; (4) energy band used in the analysis in units of keV; (5) best-fitting photon index ($^{f}$fixed); (6) best-fitting column density measured in cm$^{-2}$; (7-8) observed and absorption-corrected (intrinsic) 2--10 keV luminosities, respectively, in units of erg s$^{-1}$; (9) fit statistic results and statistical approach; and (10) spectral curvature value (\citealp{Koss16}; see Section 3.1).
   \end{tablenotes}
\end{table*}


\section{Discussion}

\begin{figure}
\centering
  \includegraphics[scale=0.45]{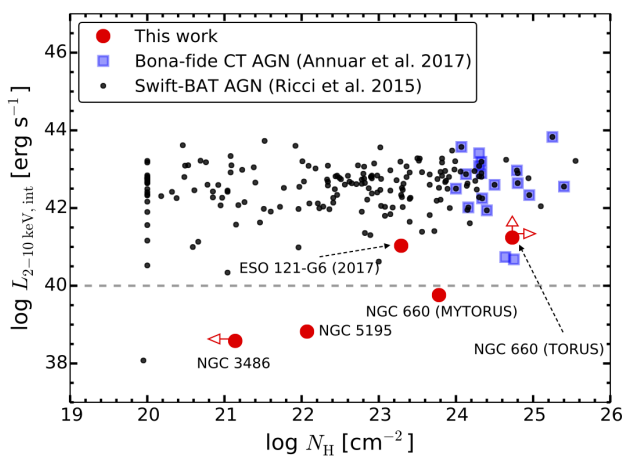}
\caption{Intrinsic 2--10 keV luminosity vs. $N_{\rm{H}}$ plot for the AGN presented in this work (red), and the \textsl{Swift}-BAT AGN located at $D \leq$ 100 Mpc (black; \citealp{Ricci15}). Dashed line indicates the luminosity threshold we use to define LLAGN. Blue squares marks the bona-fide CT AGN from \citet{Annuar17}.}
\end{figure}

In this paper, we presented broadband X-ray spectral fittings of four AGN with $L_{\rm 2-10\ keV,\ obs} \approx$ 10$^{39}$ erg s$^{-1}$ at $D \leq$ 15 Mpc to investigate the nature of their obscuration and intrinsic power. Figure 8 shows a plot of the intrinsic 2--10 keV luminosity versus $N_{\rm{H}}$ values measured for the AGN. It can be clearly seen from this figure that two of the AGN, i.e., ESO 121-G6 and NGC 660, are heavily obscured, and the remaining two, NGC 3486 and NGC 5195, appear to be unobscured and mildly obscured, respectively, and intrinsically LLAGN ($L_{\rm 2-10\ keV,\ int} <$ 10$^{40}$ erg s$^{-1}$). In this section we further investigate the results obtained from our X-ray spectral fitting with an X-ray spectral curvature analysis (Section 3.1) and undertake a joint comparison of the AGN X-ray and mid-IR data (Section 3.2). We then look at the Eddington ratios of the AGN, and discuss the difference between the heavily obscured AGN and the genuine LLAGN (Section 3.3). 

\subsection{X-ray spectral curvature analysis}

\citet{Koss16} developed a spectral curvature analysis technique which uses weighted averages of different energy bands above 10 keV to estimate the Compton-thickness of an AGN. For CT AGN, the spectral curvature (SC) value calculated is SC$_{\rm CT} \geq$ 0.4 (\citealp{Koss16}; \citealp{Baronchelli17}). We applied this technique to the two AGN in our sample that are detected at $E >$ 10 keV; i.e., ESO 121-G6 and NGC 660. For ESO 121-G6, we determined an SC value of 0.10 $\pm$ 0.02, indicating that it is not a CT AGN, in support of our X-ray spectral modelling of the AGN. For NGC 660 however, we inferred an SC value of 0.31 $\pm$ 0.13; i.e., consistent with CT absorption within the statistical uncertainties. These constraints are also consistent with our X-ray spectral modelling of the source in which we found both Compton-thin and Compton-thick solutions. On the basis of our X-ray analyses, it is therefore clear that NGC 660 is heavily obscured and may be a CT AGN. We therefore consider the $N_{\rm{H}}$ value measured from the best-fit MY{\sc{torus}} model (i.e., $N_{\rm{H}}$ $=$ 6.09$^{+3.19}_{-2.43}$ $\times$ 10$^{23}$ cm$^{-2}$) as a lower limit column density for this source. The intrinsic luminosity measured from the best-fit MY{\sc{torus}} model (i.e., $L_{\rm 2-10\ keV,\ int} =$ 5.74 $\times$ 10$^{39}$ erg s$^{-1}$) should therefore also be considered as a lower limit.


\subsection{X-ray and mid-IR data}

To further investigate the nature of the AGN in our sample, we complement the X-ray analysis of the sources with archival and new high angular resolution mid-IR data. The mid-IR continuum emission from an AGN is produced as a result of heating by the X-ray to optical (mainly ultraviolet) radiation emitted from the accretion disc. Therefore, it can be used to provide an accurate estimate for the intrinsic X-ray luminosity of the AGN, even when heavily obscured (e.g., \citealp{Gandhi09}). Given the low luminosity of these AGN, even the detection of compact mid-IR emission is important as it indicates that a dusty torus is present in these systems. However, mid-IR emission from an AGN can be contaminated by dust surrounding young forming stars. Therefore, high angular resolution observations are crucial in resolving the AGN emission from these contaminating sources. In Figure 9 and 10, we compare the intrinsic 2--10 keV luminosities measured for the AGN from our X-ray analyses, with their 12$\mu$m and [Ne{\sc{v}}] luminosities, respectively. Below we detail and discuss the mid-IR and X-ray results for each AGN. 

\begin{figure}
\centering
 \includegraphics[scale=0.45]{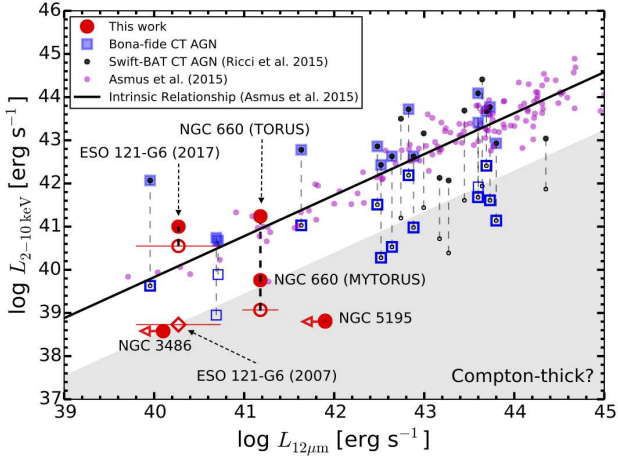}
\caption{The 2--10 keV vs. 12$\mu$m luminosities of the AGN presented in this work. The solid line corresponds to the intrinsic relationship derived by \citet{Asmus15} ($\sigma \approx$ 0.3 dex). The AGN sample used by \citet{Asmus15} to construct their relationship is plotted in magenta. Open and filled red circles indicate the observed and intrinsic 2--10 keV luminosities of our AGN as measured from X-ray spectroscopy, respectively. For CT AGN, the X-ray emission that we observed is generally attributed to X-ray photons that are scattered or reflected from the back side of the torus, which consists of just a few percent of the intrinsic power of the AGN in the 2--10 keV band (e.g. \citealp{Iwasawa97}; \citealp{Matt00}; \citealp{Balokovic14}; \citealp{Annuar17}). The grey shaded region represents a factor of 25$\times$ suppression of X-ray flux, which we have adopted as an indirect diagnostic for CT AGN (e.g., \citealp{Alexander08}; \citealp{Rovilos14}). The black circles and blue squares indicate the \textsl{Swift}-BAT CT AGN \citep{Ricci15} and bona-fide local CT AGN (adopted from \citealp{Annuar17}). The 12$\mu$m luminosities for all AGN are obtained from high angular resolution mid-IR observations with the exception of NGC 5195, for which an upper limit was obtained from {\it{WISE}}.}
\end{figure}

{\bf{ESO 121-G6}}: The intrinsic luminosity measured for ESO 121-G6 using our X-ray data is found to be consistent with that predicted by the 2-10 keV:12 $\mu$m intrinsic relationship of \citet{Asmus15} (Figure 9) and 2-10 keV:[Ne{\sc{v}}] correlation derived by \cite{Weaver10} (Figure 10). It can also be seen from Figure 9 that the observed 2--10 keV luminosity of the AGN measured from the \textsl{XMM-Newton} data in 2007 is significantly lower than our more recent X-ray data, and lies in the grey region of the plot, suggesting that the AGN may have been CT at the time of this XMM-Newton observation. The steep X-ray spectral slope measured ($\Gamma =$ 1.74$^{+1.14}_{-1.26}$) in a CT scenario would require that ESO 121-G6 was heavily CT ($N_{\rm{H}}$ $>$10$^{25}$ cm$^{-2}$) and only scattered X-ray emission was detected, which would also explain the drop in X-ray flux by $\sim$2 orders of magnitude. However, we cannot confirm this interpretation from the current data, and it is also possible that the AGN was unobscured and in a low luminosity state in 2007. Long-term monitoring with \textsl{NuSTAR} and \textsl{XMM-Newton} or \textsl{Chandra} is required to better constrain the physical origin of the variability of ESO 121-G6.

{\bf{NGC 660}}: The intrinsic luminosity we measured for NGC 660 from the best-fit MY{\sc{torus}} model is significantly lower than that estimated from its 12$\mu$m luminosity (see Figure 9). However, the intrinsic luminosity inferred using the {\sc{torus}} model, which measured a CT column density, is in agreement with that predicted by the \citet{Asmus15} relationship (see Figure 9). This is also the case when comparing the intrinsic X-ray luminosities of the AGN from the two models with its [Ne{\sc{v}}] line luminosity (see Figure 10). Therefore, considering both the mid-IR and X-ray data, we favour the CT solution from the {\sc{torus}} model to explain the X-ray emission from NGC 660. In fact, if we fixed the power-law normalisation of the MY{\sc{torus}} model so that the intrinsic luminosity of the AGN agrees with that estimated by the 12$\mu$m luminosity, we are also able to find an acceptable fit (C-stat/d.o.f $=$ 196/188), which gives a CT column density of $N_{\rm{H}}$ $=$ 5.71$^{+2.10}_{-1.06}$ $\times$ 10$^{24}$ cm$^{-2}$, consistent with the lower limit measured with the {\sc{torus}} model. Longer X-ray observations, particularly at hard X-ray energies, will be required to unambiguously confirm this.
 
{\bf{NGC 3486}}: The low X-ray luminosity, non detection by {\textsl{NuSTAR}} at $>$ 8 keV, and lack of X-ray absorption signatures suggest that NGC 3486 could be a genuine LLAGN. However, these data are also consistent with a heavily CT AGN ($N_{\rm{H}}$ $>$10$^{25}$ cm$^{-2}$) where the observed X-ray emission is scattered. Mid-IR data can help distinguish between these two competing scenarios since a heavily CT AGN would be intrinsically luminous and hence bright in the mid-IR waveband in comparison to the X-ray emission. NGC 3486 is undetected in the high-resolution mid-IR observations and the upper limit X-ray:12$\mu$m luminosity ratio places it just below the threshold for a mild CT absorption ($N_{\rm{H}}$ $\sim$ 10$^{24}$ cm$^{-2}$; see Figure 9). Consequently, we believe that the non detection of NGC 3486 in the mid-IR emission is more consistent with a LLAGN than a heavily CT AGN.

{\bf{NGC 5195:}} Unfortunately NGC 5195 lacks high spatial resolution mid-IR data, limiting the sensitivity of any mid-IR:X-ray analyses. The current upper limit on the \textsl{WISE} AGN luminosity is unable to distinguish between a Compton-thick and Compton-thin solution. However, we note that the X-ray data analyses themselves already strongly argue for a mildly obscured scenario ($N_{\rm{H}}$ $\sim$ 10$^{22}$ cm$^{-2}$).

\begin{figure}
\centering
  \includegraphics[scale=0.45]{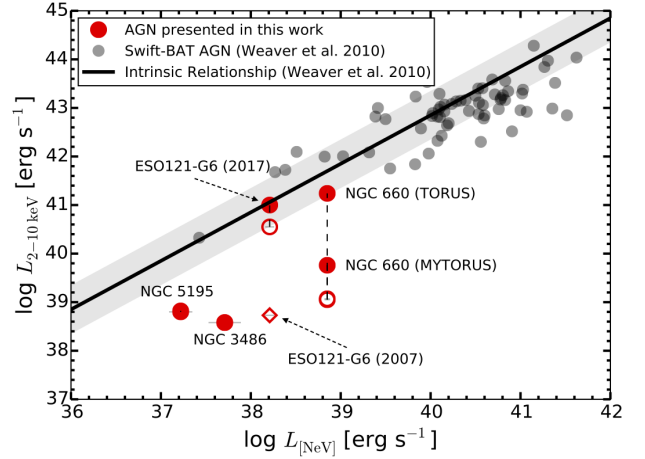}
\caption{The 2--10 keV luminosity vs. [Ne{\sc{v}}]$\lambda$14.3$\mu$m luminosity plot adapted from Figure 1 (right panel) to show the intrinsic 2--10 keV luminosity of the AGN presented in this work. Open and filled red circles indicate the observed and intrinsic 2--10 keV luminosities of our AGN as measured from X-ray spectroscopy, respectively.}
\end{figure}

The 2-10 keV luminosity from the AGN in both NGC 3486 and NGC 5195 are lower than that expected given their [Ne{\sc{v}}] luminosities (see Figure 10). While this could indicate that the AGN is intrinsically more luminous than that suggested by the X-ray luminosity, this interpretation would be inconsistent with the evidence we have presented on these two objects. An alternative scenario is that other extreme processes within the host galaxy (e.g., stellar processes and supernovae) are significantly contributing to the [Ne{\sc{v}}] line luminosity measured (\citealp{Contini97}; \citealp{Georgakakis17}; \citealp{Izotov12} and \citealp{Greene19}). This could be further investigated in the future using high angular resolution mid-IR observations, such as by the {\it{James Webb Space Telescope}} ({\it{JWST}}) Mid-IR Instrument (MIRI), which might allow us to spatially resolve the [Ne{\sc{v}}]  line emision in this galaxy and investigate its origin.

\subsection{$L_{\rm bol}$ versus $M_{\rm{BH}}$}

Figure 10 shows a plot of bolometric luminosity ($L_{\rm bol}$) versus black hole mass ($M_{\rm{BH}}$) for the AGN presented in this work. The bolometric luminosities for the AGN in our sample were estimated using the absorption-corrected 2--10 keV luminosities and assuming the bolometric correction ($\kappa$) relationship determined by \citet{Nemmen14} for LLAGN (i.e., $\kappa \approx$ 13 ($L_{\rm 2-10\ keV,\ int}$/10$^{41}$ erg s$^{-1}$)$^{-0.37}$). The black hole masses were measured using different methods; i.e., ESO 121-G6 and NGC 3486 using the bulge luminosities (\citealp{Goulding10} and \citealp{McKernan10}, respectively), whilst NGC 660 and NGC 5195 was determined using the stellar velocity dispersion (\citealp{Barth02} and \citealp{Ho09}, respectively). Based on this figure, we can see that the two AGN that were found to be heavily obscured (i.e., ESO 121-G6 and NGC 660) are accreting material at a considerably higher rate (i.e., $\lambda_{\rm Edd} = L_{\rm bol}/L_{\rm Edd} \gtrsim$ 10$^{-3}$, assuming the {\sc{torus}} model for NGC 660) than the other two unobscured intrinsically LLAGN. 




\begin{figure}
\centering
  \includegraphics[scale=0.45]{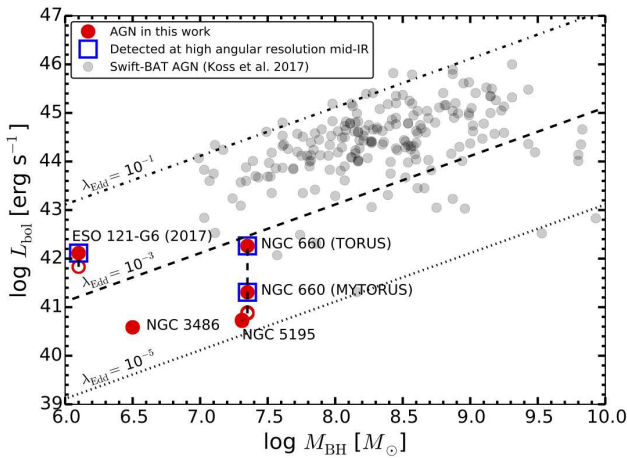}
\caption{The bolometric luminosity vs. $M_{\rm{BH}}$ for the AGN presented in this work (red; open and filled circles indicate observed and intrinsic luminosities, respectively), and the \textsl{Swift}-BAT AGN sample (grey; \citealp{Koss17}). Open blue squares mark AGN in this work which are detected at high spatial resolution mid-IR observations. The dashed-dot, dashed and dotted lines indicate constant Eddington ratios ($L_{\rm bol}$/$L_{\rm Edd}$) of $\lambda_{\rm Edd} =$ 10$^{-1}$, 10$^{-3}$ and 10$^{-5}$, respectively.}
\end{figure}


\section{Conclusion and Future Work}

In this paper, we presented \textsl{NuSTAR} observations for four AGN located at $D \leq$ 15 Mpc with observed X-ray luminosities of $L_{\rm 2-10\ keV,\ obs} \lesssim$ 10$^{39}$ erg s$^{-1}$. We combined our \textsl{NuSTAR} data with low energy data from \textsl{Chandra} (contemporaneous in most cases) or \textsl{XMM-Newton}, and performed broadband X-ray spectral fitting of the AGN to directly measure their column densities in order to determine whether they are genuinely LLAGN ($L_{\rm 2-10\ keV,\ int} <$ 10$^{40}$ erg s$^{-1}$), or deeply buried AGN. Based on the constraints from our spectral modelling, we determined that two of the AGN; i.e., ESO 121-G6 and NGC 660, are heavily obscured, and the remaining two i.e., NGC 3486 and NGC 5195, appear to be unobscured and mildly obscured, respectively, and intrinsically LLAGN. Both the heavily obscured AGN are detected at high spatial resolution in the mid-IR, indicating the presence of obscuring dust. Both the LLAGN however, lack mid-IR detection at high angular resolution (one which has been imaged in such mode but was not detected, and one has not yet been imaged in such mode). We therefore could not constraint the existence of torus for these objects at this wavelength. More sensitive observations using {\it{JWST}} for example, will be required to provide a firm constraint.

From our X-ray spectral analysis, we also found that ESO 121-G6 could be a candidate for a changing-look AGN in X-rays, and that NGC 660 is a likely CT AGN. In addition, we also suggested that NGC 3486 could be a true type 2 AGN (i.e., AGN with no BLR), as its X-ray classification (unobscured) is at odds with its optical Type 2 Seyfert classification. However, further deeper (and monitoring for ESO 121-G6) observations at different wavelengths (e.g., optical and X-rays) need to be carried out for these AGN in order to confirm these findings. 

These four AGN are a part of our nearby AGN sample located within $D \leq$ 15 Mpc. In our next paper, we will present our \textsl{NuSTAR} data and X-ray analysis results for the remaining (four) AGN in our sample which lack reliable $N_{\rm{H}}$ measurements previously. We will also combine our results for the whole $D \leq$ 15 Mpc AGN sample to form detail overview and overall census of the CT AGN population and the $N_{\rm H}$ distribution of AGN in the local universe. This will be important towards our understanding of the overall AGN population and the cosmic X-ray background radiation.

\section*{Acknowledgement}

We acknowledge financial support from the Ministry of Education Malaysia Fundamental Research Grant Scheme grant code FRGS/1/2019/STG02/UKM/02/7 (A.A.), the Science and Technology Facilities Council grant code ST/P000541/1 and ST/T000244/1 (D.M.A.), and ST/R000506/1 (P.G.). F.E.B. acknowledges support from the CONICYT grants CATA-Basal AFB-170002, FONDECYT Regular 1190818 and 1200495, and Chile's Ministry of Economy, Development, and Tourism's Millennium Science Initiative through grant IC120009, awarded to The Millennium Institute of Astrophysics, MAS. D.A. acknowledges funding through the European Union's Horizon 2020 and Innovation programme under the Marie Sklodowska-Curie grant agreement no. 793499 (DUSTDEVILS). M.B. acknowledges support from the Black Hole Initiative at Harvard University, which is funded in part by the Gordon and Betty Moore Foundation (grant GBMF8273) and in part by the John Templeton Foundation. C.R. acknowledges support from the CONICYT+PAI Convocatoria Nacional subvencion a instalacion en la academia convocatoria a\~{n}o 2017 PAI77170080, and Fondecyt Iniciacion grant 11190831. P.B. acknowledges financial support from the STFC and the Czech Science Foundation project No. 19-05599Y. E.N. acknowledges financial contribution from the agreement ASI-INAF n.2017-14-H.0 and partial support from the EU Horizon 2020 Marie Sk\l{}odowska-Curie grant agreement no. 664931.

\textsl{NuSTAR} is a project led by the California Institute of Technology (Caltech), managed by the Jet Propulsion Laboratory (JPL), and funded by the National Aeronautics and Space Administration (NASA). We thank the \textsl{NuSTAR} Operations, Software and Calibrations teams for support with these observations. This research has made use of the \textsl{NuSTAR} Data Analysis Software ({\sc{nustardas}}) jointly developed by the ASI Science Data Center (ASDC, Italy) and the California Institute of Technology (USA). The scientific results reported in this article are based on observations made by the {\it{Chandra X-ray Observatory}} and data obtained from the {\it{Chandra}} Data Archive. This research has made use of software provided by the {\it{Chandra}} X-ray Center (CXC) in the application packages CIAO. This work was also based on observations obtained with XMM-Newton, an ESA science mission with instruments and contributions directly funded by ESA Member States and NASA. Besides these, we also used observations collected at the European Organisation for Astronomical Research in the Southern Hemisphere under ESO programme 0101.B-0386(A). 

This research made use of Astropy,\footnote{http://www.astropy.org} a community-developed core Python package for Astronomy (\citealp{astropy13} and \citealp{astropy18}). We also used data obtained through the High Energy Astrophysics Science Archive Research Center (HEASARC) Online Service, provided by the NASA/Goddard Space Flight Center, and the NASA/IPAC extragalactic Database (NED) operated by JPL, Caltech under contract with NASA. 

Facilities: \textsl{Chandra}, \textsl{Gemini}, \textsl{NuSTAR}, \textsl{Swift}, \textsl{VLT}, \textsl{WISE}, \textsl{XMM-Newton}.

\section*{Data Availability}

The data used in this paper are publicly available to access and download as follow:

\begin{itemize}
 \item X-ray data
  \begin{itemize}
 \item From NASA's High Energy Astrophysics Science Archive Research Center (https://heasarc.gsfc.nasa.gov/docs/archive.html). Details of the observations, including the observation identification numbers are listed in Table 2. 
  \end{itemize}
\item Mid-IR data
 \begin{itemize}
  \item ESO121-G6: From the Gemini Observatory Archive (https://archive.gemini.edu/searchform); Program ID GS-2010B-Q-3; PI F. Bauer.
  \item NGC 660: From the European Southern Observatory Science Archive Facility (http://archive.eso.org/eso/eso\_archive\_main.html); Program ID 0101.B-0386(A); PI A. Annuar.
 \end{itemize}
\end{itemize}

\bibliographystyle{mnras}
\bibliography{annuar_nustarfaintagn_accepted}

\appendix

\section*{Affiliation}
$^{1}$ Department of Applied Physics, Faculty of Science and Technology, Universiti Kebangsaan Malaysia, 43600, Bangi, Selangor, Malaysia\\
$^{2}$ Centre for Extragalactic Astronomy, Department of Physics, Durham University, South Road, Durham, DH1 3LE, UK\\
$^{3}$ Department of Physics $\&$ Astronomy, Faculty of Physical Sciences and Engineering, University of Southampton, Southampton, SO17 1BJ, UK\\
$^{4}$ Institute of Astronomy, University of Cambridge, Madingley Road, Cambridge, CB3 0HA, UK\\
$^{5}$ European Southern Observatory, Karl-Schwarzschild str. 2, 85748 Garching bei München, Germany\\
$^{6}$ European Southern Observatory, Alonso de Cordova, Vitacura, Casilla 19001, Santiago, Chile\\
$^{7}$ Center for Astrophysics $\vert$ Harvard \& Smithsonian, 60 Garden Street, Cambridge, MA 02140, USA\\
$^{8}$ Black Hole Initiative at Harvard University, 20 Garden Street, Cambridge, MA 02140, USA\\
$^{9}$ Center for Relativistic Astrophysics, School of Physics, Georgia Institute of Technology, Atlanta, GA 30332, USA\\
$^{10}$ Instituto de Astrof{\'{\i}}sica and Centro de Astroingenier{\'{\i}}a, Facultad de F{\'{i}}sica, Pontificia Universidad Cat{\'{o}}lica de Chile, Casilla 306, Santiago 22, Chile\\
$^{11}$ Millennium Institute of Astrophysics (MAS), Nuncio Monse{\~{n}} or S{\'{\o}}tero Sanz 100, Providencia, Santiago, Chile\\
$^{12}$ Space Science Institute, 4750 Walnut Street, Suite 205, Boulder, Colorado 80301, USA\\
$^{13}$ Astronomical Institute, Academy of Sciences, Bo{\u{c}}n{\'{\i}} II 1401, CZ-14131 Prague, Czech Republic\\
$^{14}$ Department of Astronomy and Astrophysics, 525 Davey Lab, The Pennsylvania State University, University Park, PA 16802, USA\\
$^{15}$ Institute for Gravitation and the Cosmos, The Pennsylvania State University, University Park, PA 16802, USA \\
$^{16}$ Department of Physics, 104 Davey Laboratory, The Pennsylvania State University, University Park, PA 16802, USA\\
$^{17}$ Cahill Center for Astronomy and Astrophysics, California Institute of Technology, Pasadena, CA 91125, USA\\
$^{18}$ Marshall Space Flight Center, Huntsville, AL 35811, USA\\
$^{19}$ Max-Planck-Institut f{$\ddot{\rm{u}}$}r Extraterrestrische Physik (MPE), Postfach 1312, 85741, Garching, Germany\\
$^{20}$ Department of Physics and Astronomy, University of Hawaii, 2505 Correa Road, Honolulu, HI 96822, USA\\
$^{21}$  Institute for Astronomy, 2680 Woodlawn Drive, University of Hawaii, Honolulu, HI 96822, USA\\ 
$^{22}$ Eureka Scientific, 2452 Delmer Street Suite 100, Oakland, CA 94602-3017, USA\\
$^{23}$The College of New Jersey, Department of Physics, 2000 Pennington Rd, Ewing 08628, New Jersey, USA\\
$^{24}$ INAF - Osservatorio di Astrofisica e Scienza dello Spazio di Bologna, Via Piero Gobetti, 93/3, 40129, Bologna, Italy\\
$^{25}$ Department of Physics and Astronomy, Clemson University,  Kinard Lab of Physics, Clemson, SC 29634, USA\\
$^{26}$ SISSA,Via Bonomea 265, 34151 Trieste, Italy\\
$^{27}$ Dipartimento di Fisica e Astronomia, Universit\`a di Firenze, via G. Sansone 1, I-50019 Sesto Fiorentino, Firenze, Italy\\
$^{28}$ Istituto Nazionale di Astrofisica (INAF) Osservatorio Astrofisico di Arcetri, Largo Enrico Fermi 5, 50125 Firenze, Italy\\
$^{29}$N\'ucleo de Astronom\'ia de la Facultad de Ingenier\'ia, Universidad Diego Portales, Av. Ej\'ercito Libertador 441, Santiago, Chile\\
$^{30}$ Kavli Institute for Astronomy and Astrophysics, Peking University, Beijing 100871, China\\
$^{31}$ George Mason University, Department of Physics \& Astronomy, MS 3F3, 4400 University Drive, Fairfax, VA 22030, USA\\
$^{32}$ Jet Propulsion Laboratory, California Institute of Technology, 4800 Oak Grove Drive, Mail Stop 169-221, Pasadena, CA 91109, USA\\
$^{33}$ Observatorio Astronomico di Roma (INAF), via Frascati 33, 00040 Monte Porzio Catone (Roma), Italy

\label{lastpage}

\end{document}